\documentclass[twocolumn]{aastex631}
\usepackage{hyperref}

\hypersetup{hypertex=true,
colorlinks=true,
linkcolor=blue,
anchorcolor=blue,
citecolor=blue}

\newcommand\oiii{[O\,{\sc iii}]}
\newcommand\Mgii{Mg\,{\sc ii}}
\newcommand\Nii{[N\,{\sc ii}]}
\newcommand\Ciii{C\,{\sc iii]}}
\newcommand\Ciiii{C\,{\sc iv}}
\newcommand\Sii{[S\,{\sc ii}]}
\newcommand\Feii{Fe\,{\sc ii}}
\newcommand{\kms}{\ensuremath{\mathrm{km\,s^{-1}}}}
\newcommand{\ha}{\ensuremath{\mathrm{H\alpha}}}
\newcommand{\hb}{\ensuremath{\mathrm{H\beta}}}
\newcommand{\ergs}{\ensuremath{\mathrm{erg\,s^{-1}}}}

\usepackage{booktabs}
\usepackage{xcolor}
\usepackage{soul}
\begin{document}

\title{Newly Discovered Changing-look Active Galactic Nuclei from SDSS and LAMOST Surveys}

\author{Qian Dong}
\affiliation{Department of Astronomy, Xiamen University, Xiamen, Fujian 361005, People's Republic of China}

\author{Zhi-Xiang Zhang}
\affiliation{Department of Astronomy, Xiamen University, Xiamen, Fujian 361005, People's Republic of China}
\affiliation{College of Physics and Information Engineering, Quanzhou Normal University, Quanzhou, Fujian 362000, People's Republic of China}

\author{Wei-Min Gu}
\affiliation{Department of Astronomy, Xiamen University, Xiamen, Fujian 361005, People's Republic of China}

\author{Mouyuan Sun}
\affiliation{Department of Astronomy, Xiamen University, Xiamen, Fujian 361005, People's Republic of China}

\author{Yong-Gang Zheng}
\affiliation{Department of Physics, Yunnan Normal University, Kunming, 650092, People's Republic of China}

\correspondingauthor{Zhi-Xiang Zhang, Wei-Min Gu}
\email{zhangzx@xmu.edu.cn, guwm@xmu.edu.cn}

\begin{abstract}

Changing-look Active Galactic Nuclei (CL~AGNs) exhibit drastic variations in broad emission lines (BELs), the mechanism of which remains unclear. Expanding the sample of CL~AGNs is helpful to reveal the mechanism. This study aims to identify more CL~AGNs by cross-matching spectroscopic data from the Sloan Digital Sky Survey (SDSS) and the Large Sky Area Multi-Object Fiber Spectroscopic Telescope (LAMOST). Our approach to identify CL~AGNs is based on the automatic spectral fitting, followed by detailed visual inspections. We identify a sample of $51$ CL~AGNs through this method, in which $40$ CL AGNs are newly discovered. Within this sample, $41$ CL~AGNs primarily show the variability of the H$\beta$ line, nine exhibit obvious changes in both the H$\beta$ and H$\alpha$ lines, and one source mainly displays variations in the H$\alpha$ line. Our findings reveal that the sequence of appearance and disappearance of the BELs aligns with the known CL sequence. In addition, we identify 31 candidates exhibiting drastic variations in BELs without accompanying significant photometric variability. We estimate the black hole mass and Eddington ratio for all sources, which range from $2.5\times 10^6$ to $8.0\times 10^8 M_\odot$ and from $0.001$ to $0.13$, respectively. Similar to other studies, we also find that the Eddington ratios of CL~AGNs and candidates are lower than those of typical AGNs. Our results support the hypothesis that the CL behavior is driven by the state transitions of the accretion disk.

\end{abstract}

\section{Introduction}\label{sec:Introduction}

Active Galactic Nuclei (AGNs) have luminosities that can be comparable to or even far greater than their host galaxies, often exhibiting significant variability \citep{1997ARA&A..35..445U,2000ApJ...540..652W,2001sac..conf....3P,2002MNRAS.329...76H}. It is generally believed that at the center of an AGN lies a supermassive black hole (SMBH) that continuously accretes matter, producing intense radiation across the entire electromagnetic spectrum. AGNs are observationally classified into Type~1 and Type~2 based on their optical spectroscopic features. Type~1 AGNs show both broad emission lines (BELs, $1000-20000\,\mathrm{km\,s^{-1}}$) and narrow emission lines (NELs, $300-1000\,\mathrm{km\,s^{-1}}$) in their optical spectra, whereas Type~2 AGNs exhibit only NELs. According to the AGN unification model \citep{1995PASP..107..803U}, Type 2 AGNs also produce BELs; however, these BELs are obscured by a dusty torus along our line of sight. The different appearances of Type 1 and Type 2 AGNs are thus explained by our viewing angle \citep{1993ARA&A..31..473A,2006A&A...457...61R}. Spectropolarimetric observations provide evidence for this scenario. Polarized optical spectra of some Type 2 AGNs exhibit BELs, demonstrating the existence of the obscured broad-line region (BLR). However, some Type 2 AGNs show no detectable polarized BELs even in deep spectropolarimetric observations \citep{2001ApJ...554L..19T,2007ApJ...660.1072W}. Several studies argue that these latter objects are `true' Type 2 AGNs, where the absence of BELs is related to the low accretion rate rather than geometric obscuration \citep[e.g.,][]{2011ApJ...733...60T,2014Natur.513..210S,2015ARA&A..53..365N}. The accretion rate may also play a role in the classification of AGN.

Recent studies have found that BELs in AGNs can appear (turn-on) or disappear (turn-off) over several years, leading to the classification of these objects as Changing-look AGNs \citep[CL~AGNs;][]{lamassa2015discovery,2018MNRAS.480.4468R}. The first observed changing-look (CL) phenomenon was reported by \citet{tohline1976variation}, who noted a significant weakening of the \hb\ emission line in the Seyfert galaxy NGC~7603 between November 6, 1974, and November 8, 1975. \citet{2014ApJ...796..134D} collected the observations of Mrk 590 and found it changed from being classified as a Seyfert 1 galaxy to a Seyfert 2 galaxy. \citet{lamassa2015discovery} discovered the first CL quasar, which changed from Type 1 to Type 1.9 between 2000 and 2010. 

As more observational data became available, researchers employed various methods, such as optical spectra, optical variability, and mid-infrared (MIR) data, to identify more CL~AGNs. \citet{yang2018discovery} identified $21$ new CL~AGNs, in which fifteen CL~AGNs are selected from Sloan Digital Sky Survey (SDSS) and Large Sky Area Multi-Object Fiber Spectroscopic Telescope (LAMOST) observations according to their \hb\ emission line variations, and six CL~AGNs are selected from the photometric variability of W1 band of Wide-field Infrared Survey Explorer survey \citep[WISE;][]{Wright2010AJ....140.1868W,2018ApJS..234...23A}. \citet{2018ApJ...864...27S} found that WISE J105203.55+151929.5 exhibited high variability in its MIR light curve and identified it as a CL AGN through spectral observations with the Palomar telescope. \citet{sheng2020initial} found $300$ CL~AGN candidates based on significant MIR variations and color changes from AGN-like to galaxy-like. Subsequent spectral observations confirmed six as CL~AGNs. \citet{2024ApJS..270...26G} identified 130 CL~AGNs based on emission line variations in \hb, \ha, \Mgii, \Ciii, and \Ciiii, using data from the Dark Energy Spectroscopic Instrument (DESI; \citealt{2013arXiv1308.0847L,2016arXiv161100036D,2016arXiv161100037D}) survey and SDSS. To date, approximately 370 CL~AGNs have been identified in previous works \citep{macleod2016systematic, yang2018discovery, 2019ApJ...883...31F, Magnier_2020, 2022ApJ...933..180G, 2022arXiv221003928W, 2022MNRAS.511...54H,2024ApJS..270...26G, 2024ApJ...966..128W, 2024ApJ...966...85Z}

Additionally, some CL~AGNs have been observed to exhibit repeated CL phenomena. For instance, Mrk 590 was first observed in 1973 without the \hb\ BEL. In 1989, \hb\ BEL became strong, yet subsequent observations showed it was almost invisible again in 2006 \citep{denney2014typecasting}. Mrk 1018 has long-term spectroscopic monitoring and also exhibits the repeated CL phenomena \citep{2016A&A...593L...8M,2021MNRAS.506.4188L,2024arXiv241118917L}. About 17 CL~AGNs exhibit the repeat CL phenomenon \citep{marin2019changing,2025ApJ...981..129W}.

The term `CL AGN' has evolved in recent literature. \cite{2020MNRAS.491.4925G} introduced the term ‘Changing-state’ AGN (CS AGN), which exhibits variability in both spectra and light curves. The majority of the literature does not clearly distinguish between CL AGN and CS AGN \citep[e.g.,][]{2020MNRAS.498.2339R,2025ApJ...981..129W}. Throughout this work, we adopt the term `CL~AGN'. The physical origin of the CL~AGNs is still a matter of debate. Several models have been proposed to explain the CL phenomenon of AGNs, primarily falling into three categories \citep{2023NatAs...7.1282R}:
\begin{enumerate}
\item \textbf{The Dust Shadowing Effect}, a dust cloud moves in and out of the observer’s line of sight, causing changes in BELs \citep{holt1980x}. However, the dynamical timescale of the obscuration is longer than the transition timescale of most CL~AGNs \citep{lamassa2015discovery,2016MNRAS.455.1691R,2017ApJ...846L...7S}. Additionally, the obscuration effect cannot adequately explain the significant variability of MIR light curves \citep{2016MNRAS.457..389M,2016ApJ...826..188R,2016MNRAS.455.1691R}.

\item \textbf{Tidal Disruption Events (TDEs)}, stars are torn apart by the SMBH tidal forces, resulting in temporal mass accretion \citep{2015MNRAS.452...69M,padmanabhan2020changing}. The CL behavior of a few CL~AGNs may be related to the TDEs \citep[e.g.,][]{2015MNRAS.452...69M}. However, most CL~AGNs stay in bright state for long time (years or decades), which is inconsistent with the power-law decline after the peak of the TDEs \citep[e.g.,][]{2016MNRAS.455.1691R,2016ApJ...826..188R,2019ApJ...874....8M}.

\item \textbf{Variations in the accretion rate}, state transitions in accretion disks lead to the observed CL phenomenon \citep{noda2018explaining,2021ApJ...916...61F,2019ApJ...883...76R,2020A&A...641A.167S,2023ApJ...958..146W,2023NatAs...7.1282R}. This model is supported by most current studies. The physical mechanism of CS AGN is also correlated with the variation of accretion rate.
\end{enumerate}

The timescale for the transition of confirmed CL~AGNs ranges from several months to decades, which is markedly shorter than the viscous timescale predicted by the classical standard thin disk model \citep{1973A&A....24..337S}, posing a challenge to traditional accretion theories.

A comprehensive sample of CL~AGNs is crucial for statistically determining the physical mechanisms behind the CL phenomenon, which also aids other research areas, such as studies of AGN host galaxies \citep{2020MNRAS.498.3985Y,2022ApJ...933...70L,2019ApJ...876...75C,2022ApJ...926..184J}. Our work aims to identify new CL~AGNs using repeated spectroscopic data from SDSS and LAMOST, both of which provide high-quality spectral data. We use an automated program to select samples, followed by a visual inspection of the candidates. Compared to previous research, we use spectroscopic data with extended temporal coverage and perform spectra fitting to obtain the properties of the emission lines, which allows us to identify more CL AGNs. We identify 51 CL~AGNs and 31 candidates through this process.

The paper is organized as follows: Section \ref{sec:data_asys} introduces the data and spectral fitting tools used in this work. Section \ref{sec:selection} describes the procedure for selecting CL~AGNs. Sections \ref{sec:discuss} and \ref{sec:summary} provide the discussion and summary, respectively. Throughout this work, we adopt the $\mathrm{\Lambda CDM}$ cosmology with $H_0 = 67.66\,\mathrm{km\,s^{-1}\,{Mpc}^{-1}}$ and $\mathrm{\Omega_m} = 0.30966$ \citep{Planck2020}.

\section{Data and SPECTRAL ANALYSIS}\label{sec:data_asys}

\subsection{Spectroscopic data}
To identify CL~AGNs, we select AGNs or galaxies with repeated spectroscopic observations from both the SDSS and LAMOST surveys. For the SDSS data, we use the SDSS Data Release 18 (DR18) catalog, which includes objects from earlier observations \citep{almeida2023eighteenth}. We focus on data from the SDSS and Baryon Oscillation Spectroscopic Survey (BOSS) projects, which have fiber diameters of 3\arcsec\ and 2\arcsec, respectively, covering a wavelength range of 3600\,\AA\ to 10,400\,\AA\ with a typical resolution of approximately 2000 \citep{Adelman2008,Alam2017MNRAS}.

SDSS DR18 contains 5,824,700 spectroscopic observations, of which 4,882,316 are high-quality spectra (\texttt{zWarning} = 0 or 16). Among these high-quality data, 3,848,269 spectra are classified as quasars (`QSO') or galaxies (`GALAXY'). We use these classified spectra to cross-match with the LAMOST data. 

LAMOST, located at Xinglong Observatory\footnote{\url{http://www.xinglong-naoc.cn/html/en}}, is a spectroscopic survey telescope capable of simultaneously observing 4,000 spectra \citep{cui2012large,zhao2012lamost}, with a fiber diameter of 3.3\arcsec. We utilize the LAMOST low-resolution data, covering wavelengths from 3690\AA\ to 9100\AA\ with a resolution of R $\sim$ 1800. Since 2012, LAMOST has observed over 11 million low-resolution spectra. Our sample consists of data from the LAMOST DR10 and DR11 v0 catalogs, including 360,887 spectra classified as ‘QSO’ or ‘GALAXY’. We require spectra to have a signal-to-noise ratio (SNR) greater than $5$ and to be marked as normal (z $\neq$ -9999). After applying these criteria, the number of `QSO' or `GALAXY' spectra is 358,765. Due to the difference of scientific goal and survey depth, the fraction of QSO and galaxy spectra is lower in LAMOST than in SDSS.

The fiber apertures of the SDSS and LAMOST surveys are similar. For nearby AGNs, observations of the same objects by both surveys have comparable host galaxy contributions, thereby minimizing interference from differences in host contributions. The resolutions of the two surveys are also comparable, allowing a direct comparison of the spectral fitting results. The observation times of the SDSS samples range from 1998 to 2024, with most spectra observed between 2010 and 2019. The LAMOST sample was observed from 2012 to 2024. The average time interval between SDSS and LAMOST spectra is approximately 11 years. 

\subsection{Cross-matching of SDSS and LAMOST data}
We use the software \texttt{TOPCAT} \citep{2005ASPC..347...29T} for cross-matching the LAMOST catalog (DR10, DR11 v0) and SDSS DR18 to obtain the objects with repeated observations. When performing the cross-match, we set the match radius to $2$\arcsec\ and the match selection to `all matches'. We set an additional redshift criterion for the same object ($\bigtriangleup z < 0.002$, where $\bigtriangleup z$ is the redshift difference of repeated spectra). We obtain $211,271$ pairs of spectra, each pair corresponding to two repeated observations of the same source. We then perform spectral fitting on $211,271$ pairs of repeated spectra to extract the emission line properties. In each pair, the spectrum with a high \hb\ BEL flux is denoted as the high-state spectrum and vice versa. 

\subsection{Spectral fitting}\label{spectral fitting}
We use the publicly available wrapper package QSOFITMORE \citep{fu2021finding} to fit the optical spectra, which builds on the capabilities of PyQSOFit \citep{2018ascl.soft09008G}. QSOFITMORE applies the dust maps of \citet{2014A&A...571A..11P} and the interstellar extinction law of  \citet{2019ApJ...877..116W} for extinction correction. Principal Component Analysis (PCA) is employed to decompose the spectrum into galaxy and quasar components for sources with redshift less than $1.16$. The template spectra used for the PCA contain five galaxy eigenspectra and twenty quasar eigenspectra. This process identifies and subtracts the host galaxy component, yielding a pure quasar spectrum \citep{fu2021finding}.

The pure quasar spectrum is first fitted with a pseudo continuum model in line-free wavelength ranges. The pseudo continuum component includes a power-law, a \Feii\ template, and a polynomial component. The power-law component represents the continuum emission from the accretion disk. The \Feii\ emission lines are blended multiple lines, which are modeled by the optical \Feii\ templates \citep{1992ApJS...80..109B} and UV \Feii\ templates \citep{2001ApJS..134....1V,2007ApJ...662..131S,2006ApJ...650...57T}, respectively. The polynomial component is added to isolate emission lines from the continuum component with atypical shapes that are caused by dust absorption or poor flux calibration.

The continuum-subtracted spectrum contains BELs and NELs, which are fitted by multiple Gaussian functions. This study primarily focuses on the variations of the BEL components of \hb\ and \ha. The broad \hb\ component is fitted with three Gaussian functions, and the narrow component is fitted with one Gaussian function. We use the Full Width at Half Maximum (FWHM) of 1200\,\kms\ as a criterion to distinguish between broad and narrow components. The adjacent \oiii$\lambda 4959$ and \oiii$\lambda5007$ NELs are each fitted with two Gaussian profiles, representing the `wing' and `core' of the emission lines. For the \ha\ emission line, we employ three Gaussian functions to fit the broad component and one Gaussian for the narrow component. The \Nii$\lambda 6549$, \Nii$\lambda 6585$, and \Sii\ NELs around \ha\ are each fitted with a single Gaussian. Overall, our spectral fitting methodology is same to previous works
\citep[e.g.,][]{Shen_2011,fu2021finding,2021ApJ...912...20J,Jin_2023}.

A Monte Carlo (MC) approach is used to estimate the uncertainties of the spectral fitting results. In each MC trial, the spectrum is fitted after adding random Gaussian noise. The flux uncertainty is used as the noise standard deviation. 

\begin{figure*}
\plotone{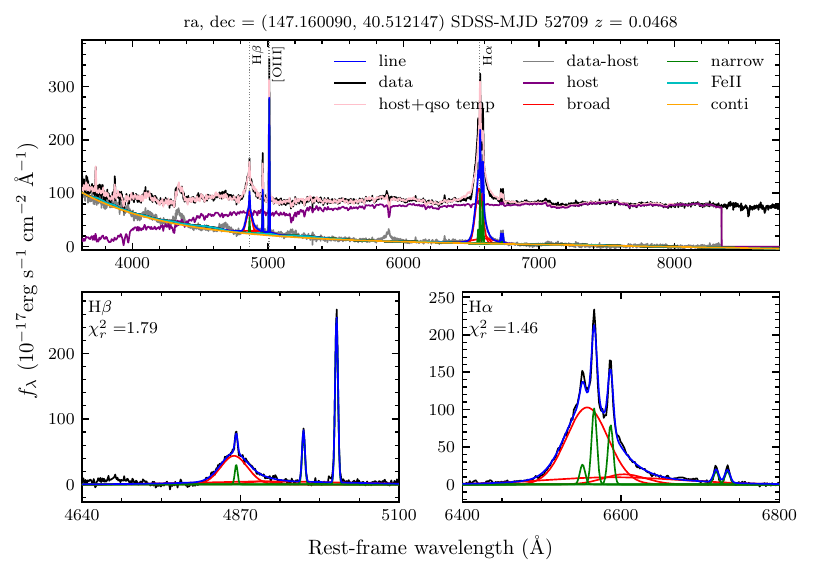}
\plotone{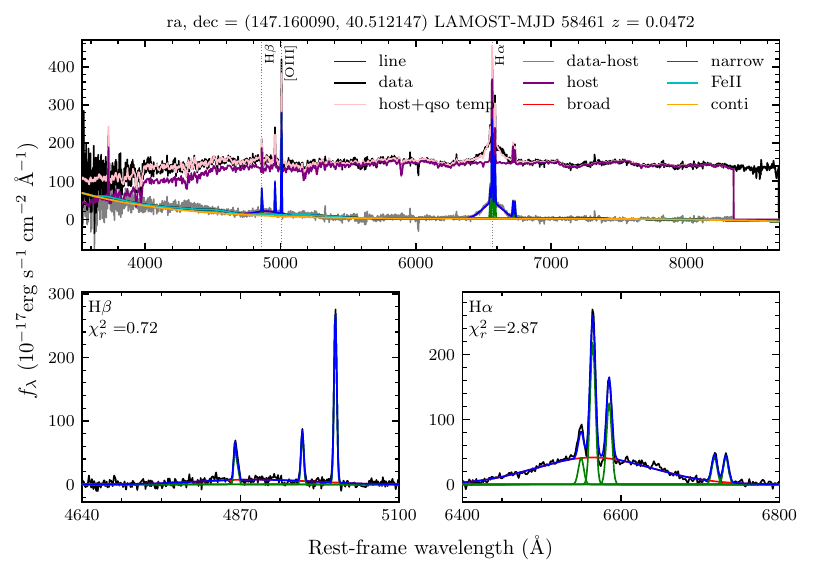}
\caption{ The spectral fitting results of the SDSS and LAMOST spectra of the CL~AGN, J094838.42+403043.7. The top panel shows the fitting result for the SDSS spectrum, where the observed spectrum (black) is decomposed into the host galaxy (purple), power-law continuum (orange), \Feii, and emission lines (blue). The emission lines include the NELs (green) and BRLs (red), which expressed as `narrow' and `broad'. The second left and right panels illustrate the details of the fitting result near the \hb\ and \ha\ emission line wavelength ranges, respectively. The third and bottom panels similarly display the fitting results for the LAMOST spectrum.  \\ (The complete figure set, including both spectral fits and light curves (82 images), is available in the online
Journal.)}
\label{fig:spectral_fitting}
\end{figure*}

Figure \ref{fig:spectral_fitting} presents an example of the spectral fitting results for two spectroscopic observations of the object J094838.42+403043.7. As seen in the figure, both spectra are well-decomposed into the host galaxy, power-law continuum, narrow, and broad emission line components. The fitting results for this example reveal a significant change in the broad component of the \hb\ emission line, transitioning from a prominent feature in the SDSS spectrum to almost completely disappearing in the LAMOST spectrum. The flux of the \ha\ emission line also shows substantial variation; however, a weaker broad \ha\ emission line component remains visible in the LAMOST spectrum. We also present the spectral fitting of an low-SNR case (see Figure \ref{fig:low_SNR} in Appendix \ref{Appendix_low_SNR}).

\subsection{Flux calibration}
The \oiii\ emission lines in AGN spectra originate from the narrow-line region (NLR), corresponding to radiative regions on the kpc scale. Therefore, their fluxes are not expected to exhibit significant variations over several decades. Consequently, the \oiii\ emission lines are widely used for calibrating the spectral flux, allowing for the comparison of flux variations of the continuum and BEL components of AGN spectra observed at different times \citep[see,][]{2013ApJ...779..109P,Fausnaugh2017}. The LAMOST spectra are only relatively flux-calibrated, lacking absolute flux calibration. To compare the flux variations of the BELs between SDSS and LAMOST spectra, we need to cross-calibrate the flux of LAMOST to SDSS. \cite{1992PASP..104..700V} indicated that the uncertainty in the flux calibration factor derived from \oiii$\lambda5007$ is below $5\%$. Therefore, we use the \oiii$\lambda5007$ emission line to do the calibration. Our specific method is as follows: first, we obtain the integrated flux of the \oiii$\lambda5007$ emission line in both SDSS and LAMOST spectra; we calculate the ratio of the SDSS \oiii$\lambda5007$ flux to the LAMOST one as the scale factor; third, we apply this scale factor to the LAMOST spectra to complete the calibration. 

\subsection{Photometric data and inter-calibration}\label{sec:photometric data}
To investigate the photometric variability, we collect multi-epoch photometric observations from several publicly available time-domain surveys, which include the Catalina Real-time Transient Survey (CRTS, \citealt{2009ApJ...696..870D}), All-Sky Automated Survey for Supernovae (ASAS-SN, \citealt{2017PASP..129j4502K}), the Palomar Transient Factory (PTF, \citealt{2009PASP..121.1395L}), Zwicky Transient Facility (ZTF, \citealt{2019PASP..131a8002B}), and the MIR observations from WISE \citep{Wright2010AJ....140.1868W}.

CRTS \citep{2009ApJ...696..870D} is a subproject of the Catalina Surveys project that began in November 2007. Its primary scientific goal is to identify variable source populations with significant amplitude variability using three telescopes: the $1.5 \, \rm{m}$ Mt.Lemmon reflector telescope, the $0.7 \, \rm{m}$ Catalina Schmidt Telescope, and the $0.5 \, \rm{m}$ Uppsala Schmidt telescope. This survey employs unfiltered CCD observations, achieving $V$-band observations of roughly 20 mag.

PTF operated from March 2009 to December 2012, utilizing the CFH12K mosaic camera installed on the $1.2 \, \rm{m}$ Samuel Oschin telescope at the Palomar Observatory in California \citep{2009PASP..121.1334R, 2009PASP..121.1395L}. This wide-field survey primarily conducts observations in two broad bands, the $g$ and $r$ bands. A successor to the PTF is an astronomical survey project aimed at discovering transient astronomical events, ZTF \citep{2019PASP..131a8003M}, which began operations in 2018, utilizing the $1.2 \, \rm{m}$ Samuel Oschin telescope with a new camera that has a field of view about $47$ square degree. This upgrade allows ZTF to scan the sky more quickly and efficiently \citep{2019PASP..131a8002B}.

ASAS-SN is a global astronomical project to cover the entire sky, which currently consists of 24 telescopes. The $V$-band observation depth of ASAS-SN reaches approximately 17 mag \citep{2017PASP..129j4502K}. 

WISE scans the entire sky covering four bands, W1, W2, W3, and W4, which correspond to wavelengths of $3.4 \, \rm{\mu m}$, $4.6 \, \rm{\mu m}$, $12 \, \rm{\mu m}$, and $22 \, \rm{\mu m}$, respectively \citep{Wright2010AJ....140.1868W}. This project began in January 2010 and completed its first full-sky scan in six months. WISE subsequently conducted continuous observations between July 2010 and February 2011, focusing on the W1 and W2 bands. After a period of dormancy, observations for these two bands were resumed in December 2013 under the Near-Earth Object WISE Reactivation mission (NEOWISE-R; see \citealt{2014ApJ...792...30M}). 
 
Given the discrepancy in the apertures and zero-points among ZTF, PTF, CRTS, and ASAS-SN, inter-calibration of these light curves is necessary. AGNs exhibit similar optical variability characteristics across different bands. The distinctions in different bands are mainly express in the `bluer-when-brighter' trend and time delay of several days \citep{2020ApJ...900...58Y,2022ApJ...929...19G}. As noted by \cite{2005ApJ...633..638W}, the difference of variability amplitude between adjacent bands is approximately $0.03\,\mathrm{mag}$. For simplicity, we assume that the variability amplitudes of different optical bands are same. We apply a scaling method (e.g., see \citealt{1991ApJ...368..119P}) to align the light curves from ASAS-SN, PTF, and CRTS with the ZTF light curve, $m' = m + C$, where $m$ and $m'$ are the magnitudes before and after the alignment, $C$ is the scaling factor. To determine the value of $C$ , we create a set of data pairs, with each pair consisting of two data points taken from the two light curves used for inter-calibration. The observation times of the two data points in each pair are separated by less than 15 days. By averaging the differences between these paired data points, we obtain the value of $C$.

\section{Sample selection} \label{sec:selection}

\subsection{Sample selection of high-state spectra }
We identify CL~AGN candidates based on the properties of the \hb\ emission line, which are automatically measured by QSOFITMORE. We find that QSOFITMORE may occasionally provide abnormal fitting results, such as spectra with very low SNR (e.g., SNR $< 10$) and abnormal combinations of the blue and red channels of LAMOST spectra. Additionally, for Type 2 spectra, QSOFITMORE sometimes fits an unusual BEL component (e.g., FWHM $>$ 13000\,\kms). To select objects with robust BELs and to reject misfitted objects, we establish the following selection criteria:

($\romannumeral1$) To ensure reliable flux calibration for LAMOST spectra, sources in which the \oiii$\lambda4959$ and \oiii$\lambda5007$ emission lines are not detected in the automatically measured process are excluded.

($\romannumeral2$) Objects with best-fit BEL FWHMs exceeding 13000\,\kms\ are excluded.
 
($\romannumeral3$) We require $F_{\mathrm{H\beta,na}}/ F_{\mathrm{H\beta,br}} < 5$ in the high-state spectrum, where $F_{\mathrm{H\beta,br}}$ and $F_{\mathrm{H\beta,na}}$ represent the peak flux densities of the broad (`br') and narrow (`na') H$\beta$ lines, respectively.

($\romannumeral4$) Objects with {$(F_{\mathrm{H\beta,br}} + F_{\mathrm{cont}}) / F_{\mathrm{cont}} > 1.2$, where $F_{\mathrm{cont}}$ is the continuum flux (`cont') at the center wavelength of the H$\beta$ emission line. This criterion requests that the broad H$\beta$ line contributes at least $20\%$ of the total observed flux in the high-state spectra. This ensures the visibility of the broad H$\beta$ line.

\subsection{Selection of objects with significant variations}
To quantify the changes of the \hb\ BEL, we simultaneously consider the relative changes in peak and integrated fluxes. We define $R_{\mathrm {p}}$ as:
\begin{equation} 
R_{\mathrm {p}} = \frac{F_{\mathrm{H\beta,br,h}} - F_{\mathrm{H\beta,br,l}} } { F_{\mathrm{H\beta,br,h}} }
\end{equation}
to address the relative variation of the peak flux, where $F_{\mathrm{H\beta,br,h}}$ and $F_{\mathrm{H\beta,br,l}}$ are the peak fluxes of the broad $\mathrm{H\beta}$ line in the high- (`h') and low-state (`l'), respectively. We also define $R_{\mathrm{s}}$ as:
\begin{equation}
R_{\mathrm {s}} = \frac{ S_{\mathrm{H\beta,br,h}} - S_{\mathrm{H\beta,br,l}}}  { S_{\mathrm{H\beta,br,h}}}
\end{equation}
to address the flux variation of the \hb, where $S_{\mathrm{H\beta,br,h}}$ and $S_{\mathrm{H\beta,br,l}}$ are the integrated fluxes of broad H$\beta$ line in the high- and low-state, respectively. Following previous studies (e.g., \citealt{2022ApJ...933..180G}, \citealt{2024ApJS..270...26G}), we consider the source to have an obvious variation of \hb\ BEL when both $R_{\mathrm{p}}$ and $R_{\mathrm{s}}$ are greater than $0.4$. There are 2,681 group observations with obvious changes in the \hb\ emission line. Both positive and negative changes are considered in this analysis.
 
\subsection{Visual inspection and selected objects}
\label{sec:visual_ins}
\begin{deluxetable*}{cccB}
\tablewidth{0pt}
\tablecaption{The selection of CL~AGNs and candidates from SDSS and LAMOST surveys \label{table1}}
\centering
\tablehead{\colhead{Note} & \colhead{Selection} & \colhead{LAMOST} }
\startdata
LAMOST spectral classified as Galaxy or QSO & Galaxy or QSO, $z \neq -9999$ & 358765 \\
Repeated observation with SDSS &  The cross-match radius $<2{\arcsec}$ , $\bigtriangleup z<0.002$  & 230823 \\
 Spectra containing \oiii\ emission lines & $z < 0.8$ & 211271 \\
Good spectral fits     & The selection criteria in Section 3.1  & 15830   \\
 Evident changes of H$\beta$  &  $R_{\mathrm{p}}> 0.4$ and $R_{\mathrm{s}} > 0.4$  & 2681 \\
Visual inspection &  Appearance or disappearance of H$\beta$ &  51 CL AGNs \\   
  & Without significant variability in light curves &  31 Candidates \\
\enddata
\end{deluxetable*}

After the spectral fitting and automatic selection, we manually inspect the remaining 2,681 objects. The main purpose of visual inspection is to discard objects with fitting errors and to identify those that exhibit the appearance or disappearance of BELs. During the visual inspection, we require the high-state spectra to show significant BEL features, excluding objects with poor SNR (SNR $<$ 10). For low-state spectra, we require clear visibility of \oiii\ emission lines and the NELs of \hb. We exclude sources that satisfy the selection criteria but exhibit false features or unreliable measurements in the spectra due to cosmic-ray or low SNR. After applying these criteria, we retain a sample of 88 sources.

Among these 88 sources, 44 sources have multi-epoch spectroscopic observations from SDSS and LAMOST. We visually inspect the multi-epoch spectra and discover the discrepancies between the spectra and optical light curves in six sources (J014417.28+314003.3, J084927.37+324852.8, J102025.83+413342.4, J111142.95+281126.6, J120510.23+333532.4, and J165718.41+341637.0). As shown in Figure \ref{fig_multi_no}, J102025.83+413342.4 displays obvious \hb\ BELs in both SDSS (MJD 53002 ans MJD 57140) and LAMOST spectra (MJD 55955 and MJD 56358). The subsequent LAMOST spectrum at MJD 58555 exhibits the disappearance of the \hb\ BEL, while the source reaches its brightest photometric state. This abnormal may be caused by the fiber drop during spectroscopic acquisition \citep{2016AJ....151...24A}. The NLR extends from $~10\,\rm{pc}$ to $~1\,\rm{kpc}$, which is farther to the center region than BLR. If the fiber is misaligned with the galactic nucleus, it may fail to capture photons from the compact BLR, but it can also collect photons from the NLR. This results in the absence of both broad \hb\ and \ha\ emission lines in the spectra. Consequently, these six sources are excluded from our final sample.

\begin{figure*}
  \begin{center}
    \resizebox{110mm}{75mm} {\includegraphics *{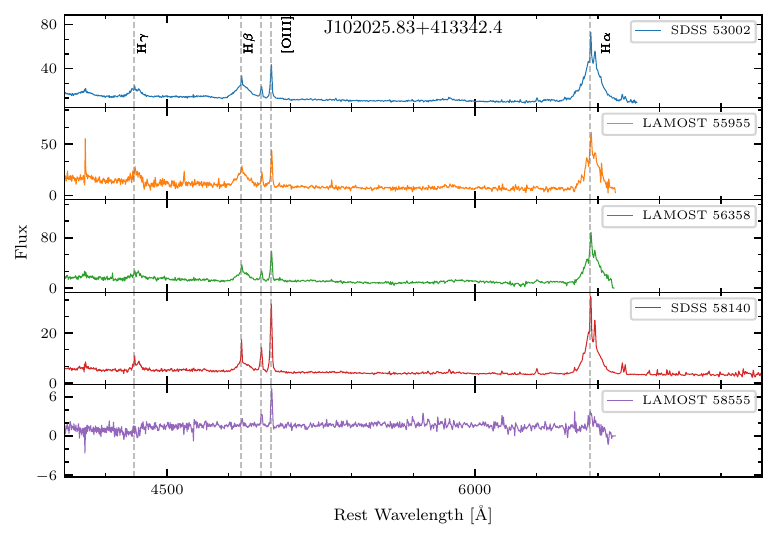}}
    \resizebox{60mm}{75mm} {\includegraphics *{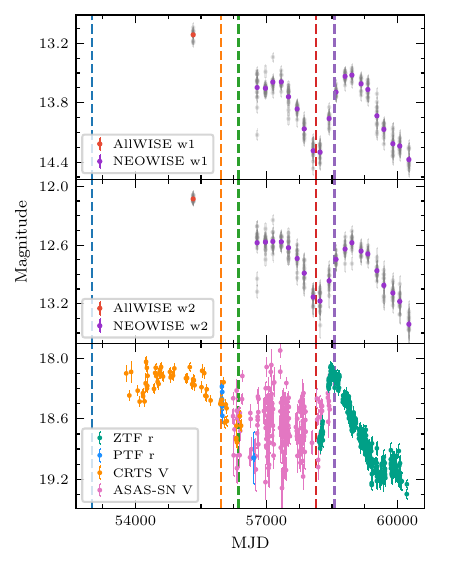}}
\caption{The spectra and light curves of J102025.83+413342.4. The left panel shows the SDSS and LAMOST spectra, while the right panel presents the light curves. The flux density of the SDSS spectrum is in units of $10^{-17}\ \mathrm{erg\ cm^{-2}\ s^{-1}\ \text{\AA}^{-1}}$, whereas the LAMOST spectrum is presented in relative flux. The five short dotted lines in different colors in the right panel indicate the epochs of the spectroscopic observations. }
\label{fig_multi_no}
\end{center}
\end{figure*}

We visually inspect the remaining 82 sources and discover some exhibiting the disappearance or emergence of both broad \hb\ and \ha\ lines, while their photometric light curves show no significant variability between spectroscopic epochs. Considering the possibility of fiber drop, we conservatively classify these sources as CL AGN candidates rather than confirmed CL AGNs. Sources that exhibit the disappearance or emergence of broad \hb\ emission line, along with either detectable broad \ha\ emission line or significant photometric variability, are identified as CL AGNs.
For the 82 targets, 51 sources are classified as CL AGNs and 31 are classified as candidates. Among the 31 candidates, two objects (J115227.49+320959.5 and J123359.13+084211.6) have confirmed as CL AGNs, which are labeled in Table \ref{CLAGNs} (in Appendix \ref{Appendix_table}). This suggests that the candidates are likely CL AGNs. 

The entire selection process of CL~AGNs and candidates  is detailed in Table \ref{table1}. Figure \ref{fig:RsRp} shows the $R_{\mathrm{p}}$-$R_{\mathrm{s}}$ distribution of the selected objects after each step of the selection process. For CL AGNs, most of them exhibit CL behaviors of the broad \hb\ line, in which nine CL~AGNs show notable changes in the \hb\ and \ha\ emission lines. One object whose \ha\, emission line show significant variations, while \hb\ emission lines are always observed as narrow lines.

\begin{figure}
\centering  
\includegraphics[width=0.48\textwidth]{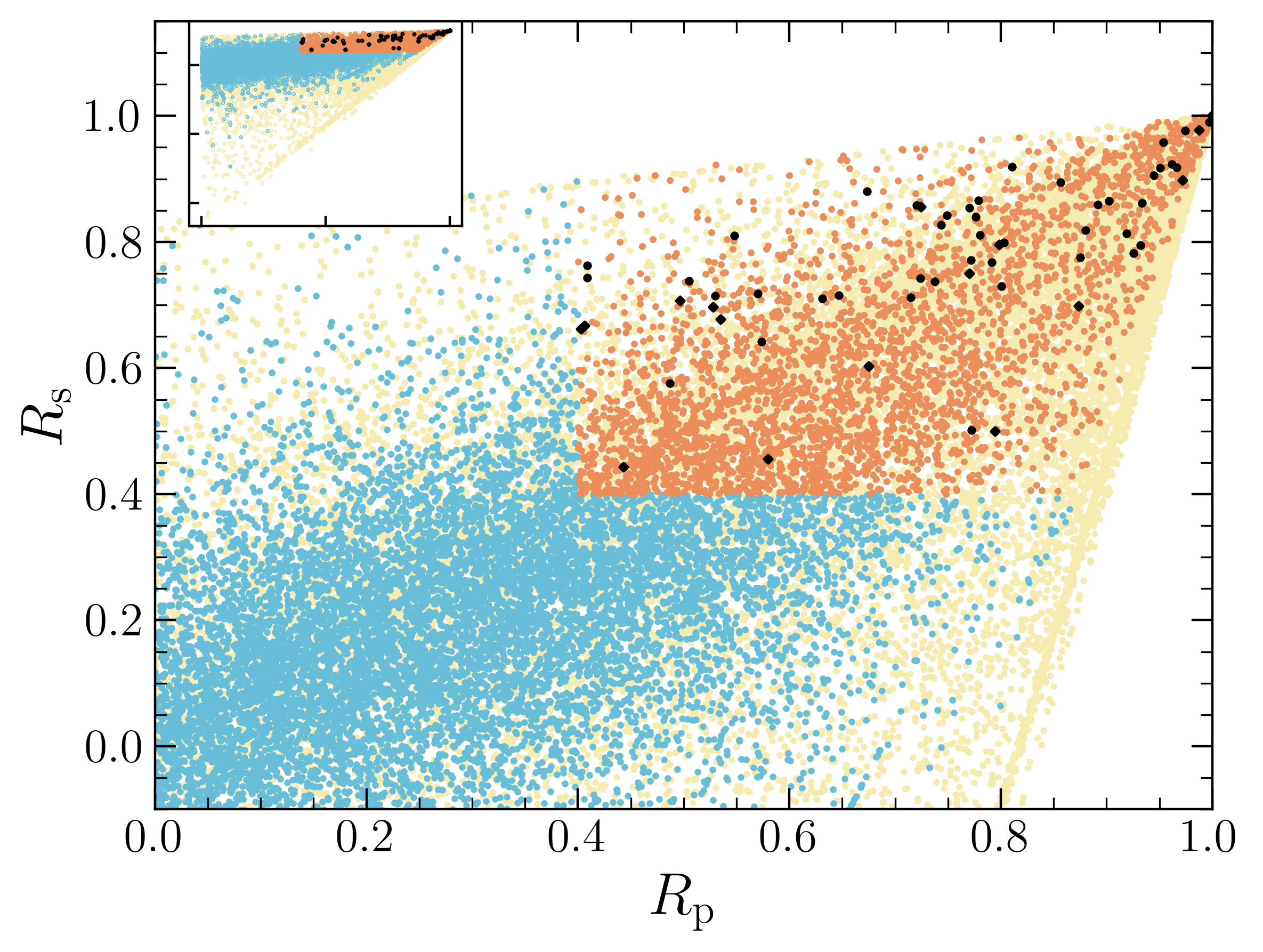}
\caption{Distribution of samples on the $R_{\mathrm{s}}-R_{\mathrm{p}}$ diagram. The small diagram at the top left represents the distribution of $R_\mathrm{s}$ and $R_\mathrm{p}$ for all samples, while the main diagram shows the distribution of samples restricted to the range of 0 to 1. The repeated observations of SDSS and LAMOST are represented by yellow points. The spectra with good fitting are represented by blue points, and the spectra of samples selected for visual inspection are represented in orange points. The black points and diamonds represent the selected CL~AGNs and candidates.}
\label{fig:RsRp}
\end{figure}

\section{Discussion}\label{sec:discuss}

\subsection{CL~AGN sample}\label{sec:dis_clsample}
As described in Section \ref{sec:visual_ins}, we select 51 CL~AGNs and 31 candidates. We cross-match our sample with the literature and find that two candidates and eleven CL AGNs had already been reported by other works \citep[see][]{2016ApJ...821...33R,yang2018discovery,2019ApJ...887...15W,2022ApJ...933..180G,2024ApJ...966..128W}. We also perform the cross-matching with the CL~AGN catalog of \cite{2024ApJ...966...85Z} and find no overlapping sources. The reported objects are labeled in Table \ref{tab:example} (in Appendix \ref{Appendix_table}).

Among the 51 CL~AGNs, seven CL~AGNs show turn-on behavior while 44 CL~AGNs display turn-off behavior. The number of turn-off type CL~AGNs far exceeds that of turn-on CL~AGNs. A similar phenomenon has also been observed in other studies (e.g., \citealt{yang2018discovery}, \citealt{2022ApJ...933..180G}). Ideally, for a complete sample, the number of turn-on and turn-off CL~AGNs should be nearly equal to ensure that the ratio of Type 1 and Type 2 AGNs remains stable. A possible explanation for the observed ratio of turn-on CL~AGNs to turn-off ones is that most of the SDSS's AGN spectra are Type 1. As a result, most of the CL~AGNs selected from the SDSS samples are of the turn-off type. A similar explanation can be found in \citet{2019ApJ...874....8M}.

\subsection{CL sequence}

\begin{figure*}
  \begin{center}
    \resizebox{110mm}{75mm} {\includegraphics *{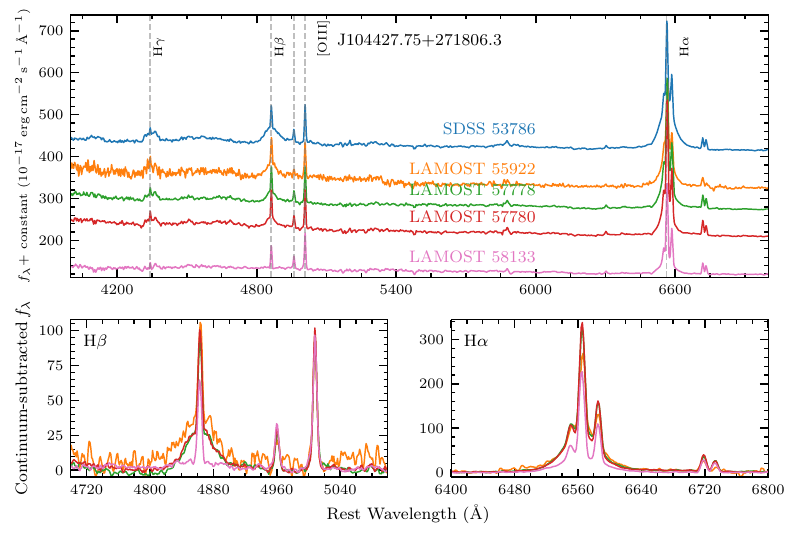}}
    \resizebox{60mm}{75mm} {\includegraphics *{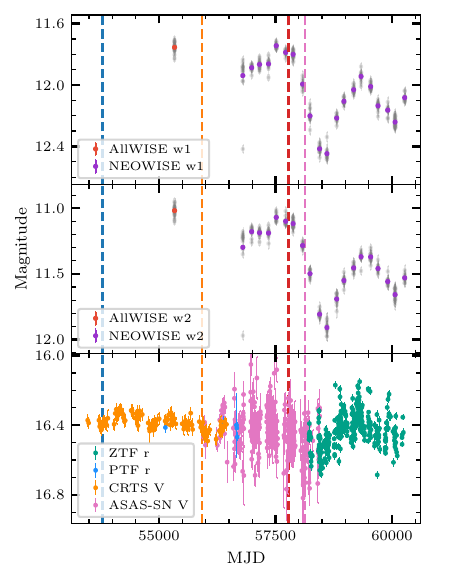}}
\caption{Spectroscopic evolution of the CL AGN J104427.75+271806.3 since 2006. The LAMOST spectra are flux-calibrated based on the SDSS spectrum obtained at MJD 53786. The left panel presents the multi-epoch spectra, \hb, and \ha\ emission line profiles. The right panel shows the light curves, with colored dotted lines indicating the epochs of spectral observations. }
\label{fig_evolution}
\end{center}
\end{figure*}

Based on multi-epoch spectroscopic observations, we identify four sources (J104427.75+271806.3, J123020.67+462745.7, J124554.71+460418.7, and J114154.47+063509.6) exhibiting a gradual decline in \hb\ emission line flux. As shown in Figure \ref{fig_evolution}, the CL AGN J104427.75+271806.3 has five spectra between MJD 53786 and MJD 58133. The initial spectra obtained at MJD 53786 and MJD 55922 exhibit prominent broad \hb\ and \ha\ emission lines. Subsequent observations at MJD 57778 and 57780 exhibit significant decline of broad \hb\ line flux compared to the initial epoch, followed by its disappearance at MJD 58133. Notably, the \hb\ broad component exhibits complete disappearance,  while the \ha\ emission line remains detectable. The five-epoch spectroscopic observations indicate the transition order of \hb\ and \ha: the broad \hb\ emission line disappears first, followed by the \ha\ emission line.   

We also examine the CL types of \hb\ and \ha\ emission lines. Of the 51 CL~AGN samples, the CL behaviors of 41 objects primarily occur in the \hb\ emission lines, and the broad \ha\ emission lines are consistently present, as indicated by `H$\beta$' in the `Line' column of Table \ref{CLAGNs} (in Appendix \ref{Appendix_table}). There are nine CL~AGNs whose H$\beta$ and H$\alpha$ emission lines both display notable changes, denoted by `H$\beta$, H$\alpha$'. The CL behavior of only one CL~AGN primarily is shown in the variations of the \ha\ emission lines, while the broad \hb\ emission line is always invisible in the two observations.

The appearance and disappearance of the broad \hb\ and \ha\ emission lines indicate that during the transition of CL~AGNs from a fully Type 2 to a Type 1 state, the broad \ha\ emission line appears first, followed by the broad \hb\ component. Conversely, during the transition from a fully Type 1 to a Type 2 state, the broad \hb\ component disappears first, followed by the broad \ha\ component. This phenomenon known as the emission line sequence, has been identified in previous studies \citep{guohx2019}.

\subsection{Timescale}
\begin{figure}
\centering
\includegraphics[width=0.48\textwidth]{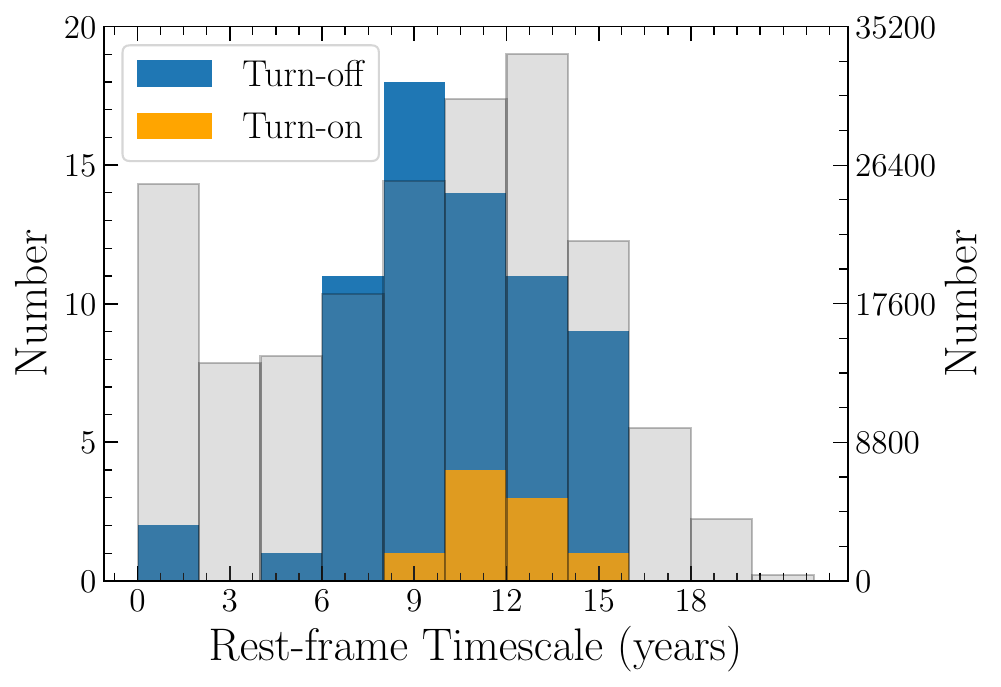}
\caption{Distributions of time intervals in the rest-frame for turn-on (orange) and turn-off (blue) sources. Gray distributions represent the time intervals between two spectroscopic observations for 211,271 objects.}
\label{Timescale_on_off}
\end{figure}

Table \ref{CLAGNs} (in Appendix \ref{Appendix_table}) lists the observation time of each CL~AGN and candidate by SDSS and LAMOST surveys. Due to there are two spectroscopic observations for most objects, it's difficult to determine the exact CL transition timescale. The time intervals of spectroscopic observations only give the upper limits of the transition timescale. Figure \ref{Timescale_on_off} illustrates the distributions of the upper limits of the turn-on and turn-off transition in the rest frame. The timescale distributions shown in Figure \ref{Timescale_on_off} are based on a total of 82 sources, without distinguishing between confirmed CL~AGNs and candidates. The orange and blue histograms correspond to turn-on and turn-off sources, respectively, and indicate that most transitions have upper timescale limits of approximately 10 years. The gray histograms represent the time intervals between two spectroscopic observations for 211,271 objects with redshifts less than 0.8 (see Table \ref{table1}). These gray histograms show that the majority of sources have spectroscopic baselines of about 10 years. The temporal distributions of the turn-on and turn-off sources are broadly consistent with that of the general spectroscopic sample. This indicates that the distribution of transition timescale upper limits is likely dominated by the sampling interval. Previous studies (e.g., \citealt{yang2018discovery,2024ApJS..270...26G,2024ApJ...966..128W,2024ApJ...966...85Z}) have reported similar distributions of transition timescale upper limits, possibly due to same reasons.

\subsection{Black hole mass and Eddington ratio}
\begin{figure*}
\centering
\includegraphics[width=0.6\textwidth]{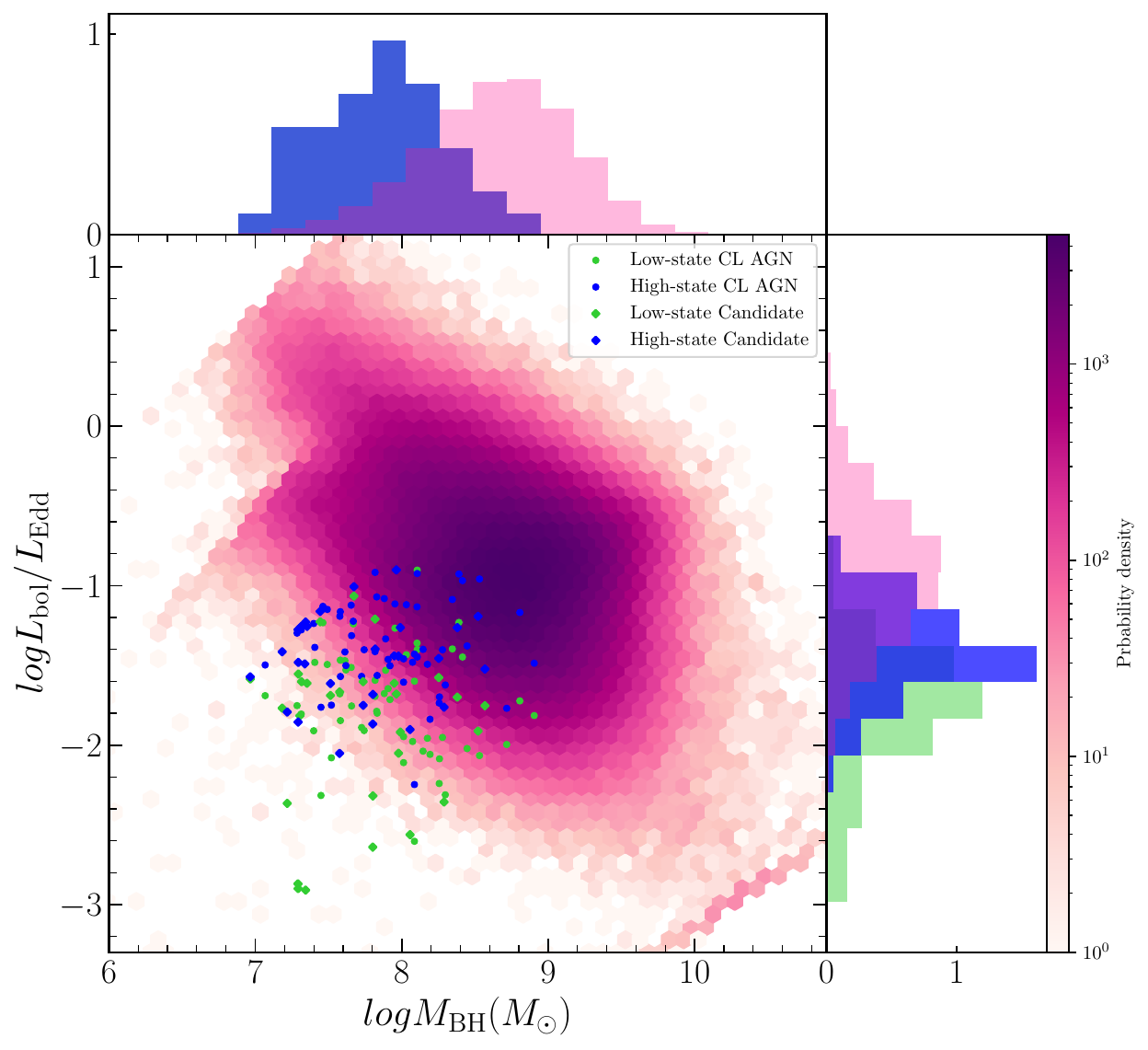}
\caption{The distribution of CL~AGNs, candidates, and typical AGNs in $\log L_\mathrm{bol}/L_\mathrm{Edd} - \log M_\mathrm{BH}$ space. The blue dots and green dots are the high-state CL~AGNs with broad \hb\ line and low-state CL~AGNs without broad \hb\ line, respectively. Candidates are represented by diamonds. The typical AGN sample is from the SDSS DR16 quasar catalog provided by \citet{2022ApJS..263...42W}. The top and right panels show the black hole mass and Eddington rate distributions, respectively. On the top panel, the blue bars represent the CL~AGNs and candidates, and the pink bars show the typical AGNs. On the right panel, the blue and green bars represent the Eddington rate in the high-state and low-state, and the pink bars show the Eddington rate of the typical AGNs.}
\label{BH_Edd}
\end{figure*}
Assuming the BLR is virialized, the mass of an SMBH can be estimated via
\begin{equation}
    M_\mathrm{BH} = f R_\mathrm{BLR} (\Delta V)^2 / G,
\end{equation}
where $f$ is a dimensionless scaling factor, $R_\mathrm{BLR}$ is the effective radius of the BLR for a given BEL, $\Delta V$ is FWHM of a BEL, and $G$ is the gravitational constant. Following the correlation between the \ha\ emission line luminosity ($L_\ha$) and the monochromatic luminosity at 5100\AA\ (\citealt{2005ApJ...630..122G}),
\begin{equation}
    \lambda L_\lambda(5100\mathrm{\AA}) = 2.4 \times 10^{43}{L_{42}}^{0.86}\,\ergs,
\end{equation}
where $L_{42} = L_\ha / 10^{42}\,\ergs$, and the $R_\mathrm{BLR} - L_{5100}$ relation \citep{2006ApJ...644..133B}, \citet{2007ApJ...670...92G} derived the black hole mass estimator as follows,
\begin{equation} 
M_{\mathrm{BH}} = 3 \times 10^{6} {L_{42}}^{0.45} (\frac{\Delta V}{\mathrm{10^3\,km\,s^{-1}}})^{2.06} \, M_{\odot}.
\end{equation}

The \ha\ BEL is typically stronger than the \hb\ BEL and provides more reliable integrated flux and width measurements. We use the \ha\ BEL properties of the high-state spectrum to estimate the black hole mass. The $\mathrm{FWHM}_\ha$ and the integrated flux of \ha\ are derived from spectral fitting results (see Section \ref{spectral fitting}). We do not directly use the fitting flux of the power-law component, as most of our sample is dominated by the host component, and the decompositions of the power-law and host components have large uncertainties. For each source, we use the \texttt{cosmology} module in \texttt{Astropy} package introduced in Section \ref{sec:Introduction} and the redshift to calculate the luminosity distance. The black hole mass ranges are from $2.54\times 10^6\,M_{\odot}$ to $8.03\times 10^8\,M_\odot$.

The Eddington accretion ratio is crucial for understanding the physics of CL~AGNs. Given the monochromatic luminosity, $L_{5100}$, and the black hole mass, we calculate the Eddington ratio as follows:
\begin{equation}
    \lambda_{\rm Edd} = \frac{L_{\rm bol}}{L_{\rm Edd}} = \frac{\mathrm{BC_{5100}} \lambda L_\mathrm{5100}}{L_\mathrm{Edd}},
\end{equation}
where $L_\mathrm{bol}$ is the bolometric luminosity, $L_{\mathrm{Edd}} \, = \,1.26 \times 10^{38} \times (M_{\mathrm{BH}}/M_{\mathrm{\odot}}) \, \mathrm{erg \, s^{-1}} $ is the Eddington luminosity, and $\mathrm{BC}_{5100}$ is the bolometric correction factor at 5100\AA; here, we adopt $\mathrm{BC}_{5100} = 9$ \citep{2000ApJ...533..631K}. 

The black hole mass and Eddington ratio of CL~AGNs and candidates are summarized in Table \ref{CLAGNs} (in Appendix \ref{Appendix_table}). Note that we excluded the object J130339.71+190120.9, who has an unreliable \ha\ fit due to the abnormal combination of the blue and red channel spectra. Figure \ref{BH_Edd} illustrates the distribution of CL~AGNs and candidates on the $\lambda_\mathrm{Edd}-M_\mathrm{BH}$ plane. The blue and green dots represent the CL~AGNs in the high- and low-states, respectively. The blue and green diamonds represent the high- and low-state candidates.
For comparison, the distribution of black hole mass and $\lambda_\mathrm{Edd}$ from the quasar sample in \citet{2022ApJS..263...42W} is also shown. 

Previous studies have extensively discussed the Eddington ratios of CL~AGNs. \citet{2019ApJ...874....8M} and \citet{2022ApJ...933..180G} found that the Eddington ratios of CL~AGNs are lower than those of the normal quasar control sample. \citet{2022APJ...927...2} revealed that the Eddington ratios of CL~AGNs are between those of QSO and low-luminosity AGNs. Additionally, \citet{2024ApJ...966..128W} further demonstrated that transitions of CL~AGNs may occur at the Eddington ratio lower than $0.01$. For our CL~AGNs and candidates, the Eddington ratio of high-state sources ranges from 0.0048 to 0.1253, while those of low-state sources range from 0.0012 to 0.1250. We find that, in Figure \ref{BH_Edd}, the Eddington ratio of our CL~AGNs and candidates are located in the lower left region compared to typical AGNs, showing that the accretion rates of CL~AGNs are lower than those of most AGNs. Our result is consistent with previous studies \citep{2019ApJ...874....8M,2021ApJ...912...20J,2022ApJ...933..180G,2022APJ...927...2,2024ApJ...966..128W}. \citet{2019ApJ...883...76R} found that CL~AGNs exhibit an inversion in the correlation between the UV-to-X-ray spectral index ($\alpha_{\mathrm{OX}}$) and the Eddington ratio at a critical Eddington ratio of $\sim 10^{-2}$. This behavior is similar to the transition of the accretion state in X-ray binaries. The Eddington ratios of our CL~AGNs are close to this critical value, supporting the hypothesis that the CL phenomenon may be driven by transitions of accretion states \citep{2017ApJ...835..144G,2020MNRAS.491.4925G}.

\subsection{BPT diagram}
\begin{figure}
\centering
\includegraphics[width= 0.48\textwidth]{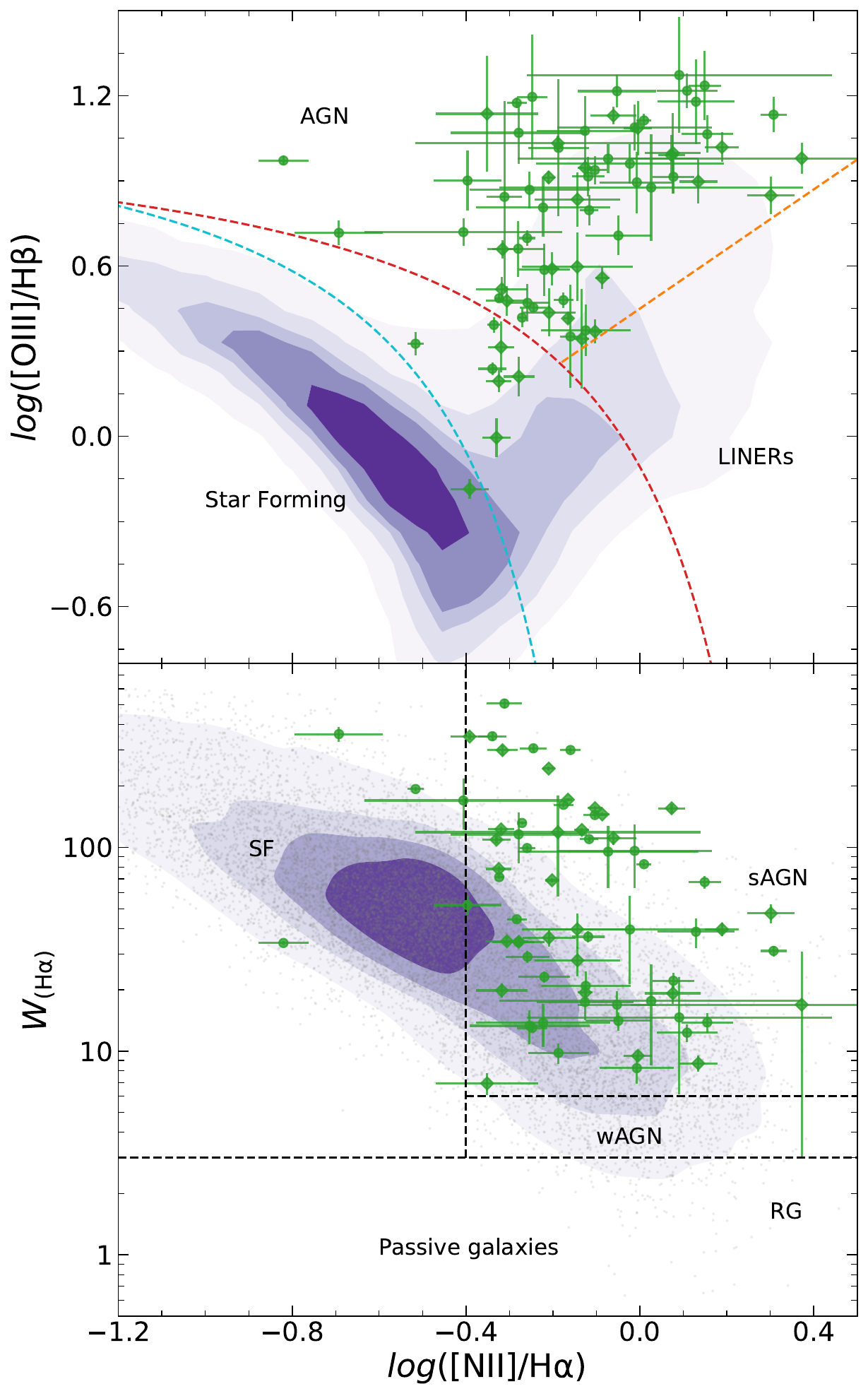}
\caption{BPT and WHAN diagram for 51 CL~AGNs (green circles) and 31 candidates (green diamonds) in our sample, and galaxies from SDSS DR12 (purple background). The blue, red, and orange dashed lines denote classification boundaries established by \citet{2003MNRAS.346.1055K}, \citet{2001ApJ...556..121K}, and \citet{2010MNRAS.403.1036C}, respectively. The WHAN diagram is divided into four parts according to \cite{2011MNRAS.413.1687C}. }
\label{BPT}
\end{figure}

In Section \ref{spectral fitting}, we derive the properties of NELs by performing spectral fitting, focusing on features linked to the photoionization of narrow-line region gas. The potential ionizing sources for these NELs include the AGN continuum, star formation, or a combination of both. To distinguish among these sources, we employ the Baldwin-Phillips-Terlevich \citep[BPT;][]{1981PASP...93....5B} diagnostic diagram, which classifies ionization origins by the flux ratio of \Nii/\ha\ to that of \oiii/\hb. Additionally, \cite{2011MNRAS.413.1687C} adopted the flux ratio \Nii/\ha\ and equivalent width of H$\alpha$ (WHAN diagram) to classify galaxies.

We chose SDSS spectra for this analysis due to their higher SNR, which enables to obtain more reliable fitting results. We plot the flux ratios of 82 sources on the BPT and WHAN diagrams (Figure \ref{BPT}). Note that we exclude eleven sources whose flux ratio uncertainties exceed 0.4. We also plot the BPT and WHAN distribution of the galaxies from SDSS DR12, using the emission line flux measurements obtained from \citet{2013MNRAS.431.1383T}. Following the classification schemes of \cite{2003MNRAS.346.1055K}, \cite{2001ApJ...556..121K}, and \cite{2010MNRAS.403.1036C}, the BPT diagram is divided into four regions: star-forming (SF), AGN-dominated, low-ionization nuclear emission-line regions (LINERs), and composite (SF and AGN) regions. 63 sources fall within the AGN-dominated region of the BPT diagram, indicating that their narrow-line emission is mainly powered by the AGN continuum. The WHAN diagram is divided into SF, weak AGN (wAGN), strong AGN (sAGN), retired galaxies (RG) and passive galaxies (PG) \citep{2011MNRAS.413.1687C}. 67 sources fall within the sAGN region of the WHAN diagram, indicating that the narrow-line emission is primarily driven by the AGN continuum. These sources exhibit consistent classifications in both the BPT and WHAN diagrams, suggesting that AGN activity is the dominant factor in shaping their emission-line properties.

The distribution in the AGN region implies that, on timescales of thousands to tens of thousands of years, these sources have consistently maintained AGN-level activity, rather than that they have been recently `switched on'. This suggests a relatively stable active phase, during which the AGN continuum has effectively ionized the surrounding NLR gas over extended periods.

\subsection{Multi-band variability behavior}

\begin{figure*}
\centering
\includegraphics[width=0.85\textwidth]{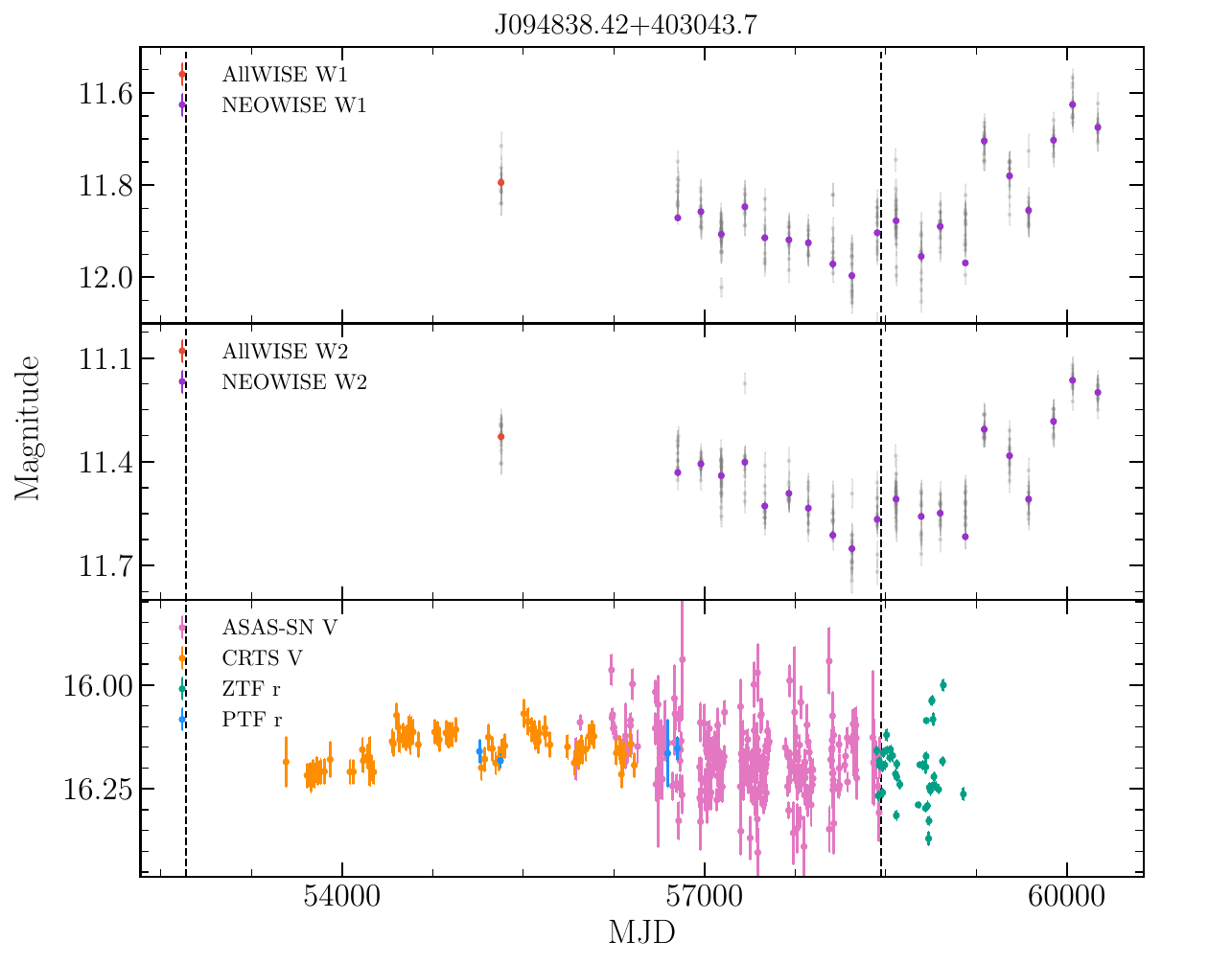}
\caption{The light curves of the CL~AGN J094838.42+403043.7. In the top and middle panels, the purple and red points represent the data collected from the NEOWISE and ALLWISE. The bottom panel shows the light curves from ZTF (green), PTF (sky blue), CRTS (orange), and ASAS-SN (pink). (Light curves for all the sources can be found in the online figure set associated with Figure \ref{fig:spectral_fitting}.)}
\label{lightcurve}
\end{figure*}

\begin{figure}
\centering
\includegraphics[width= 0.48\textwidth]{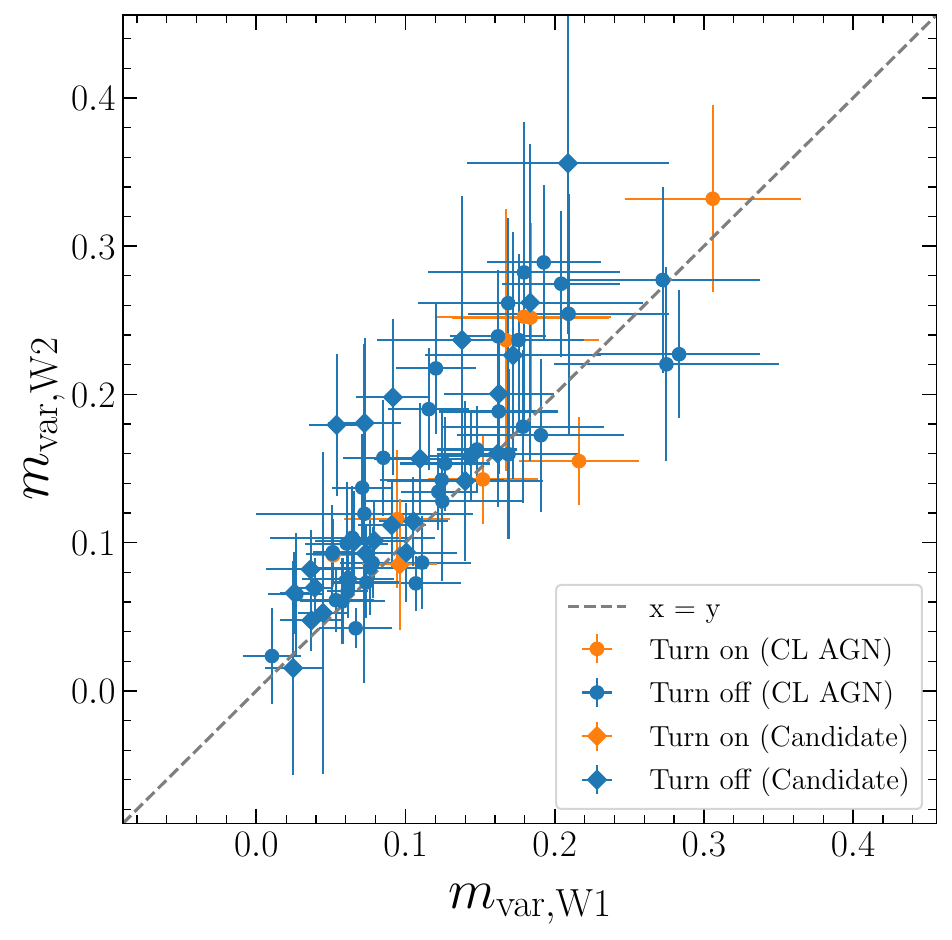}
\caption{The distribution of the variability amplitude in W1 and W2 bands for CL~AGNs and candidates. Blue circles represent turn-off CL~AGNs, while orange circles represent turn-on CL~AGNs. Candidates are represented by diamonds. The gray dashed line represents the one-to-one relation.}
\label{w1-w2}
\end{figure}

We collect photometric data for the 82 sources, including optical data from CRTS, ASAS-SN, and PTF/ZTF, as well as MIR data from WISE. For the photometric data, we perform the following data filtering and processing steps.

Firstly, we remove the data points affected by observational issues.
\begin{itemize}
 \item We exclude blended observations ($\text{Blend} \neq 0$) from the CRTS dataset to ensure the reliability of the data points.
 \item For ASAS-SN data, we remove data points with magnitude errors larger than $0.4$ or the `Quality' flag of `B'. 
 \item We remove ZTF data points affected by the moon or clouds ($\text{catflags} = 32768$). 
 \item The PTF data points suffering from the transient sources ($\text{transient\,flag} = 1$) or missing limit magnitude ($\text{limitmag = null}$) are also removed. 
\end{itemize} 

To exclude outliers, we smooth light curves using the LOWESS function \citep{doi:10.1080/01621459.1979.10481038} and calculate the residuals between the observed data and the smoothed curves. Data points with residuals exceeding three standard deviations are removed.

For the optical data, we calibrate all light curves to the ZTF-r band using a inter-calibration method described in Section \ref{sec:photometric data}. The WISE survey has approximately 12 data points within $36\,\text{hr}$. WISE conducts observations of the same sky region approximately every six months, allowing for the detection of MIR variability of objects. We combine consecutive observations within each monitoring window into a single data point, which averages successive exposures to improve the SNR, follows the method outlined in \citet{2023ApJ...950..122L}. Figure \ref{lightcurve} presents the optical and WISE light curves of J094838.42+403043.7. 
 
To further quantify the intrinsic variability, we calculate the $m_{\mathrm{var}}$ \citep{2002ApJ...568..610E} as follows:
\begin{equation}
m_{\rm var} = \sqrt{S^2-<\sigma^2_{\rm t}>},
\end{equation}
where $S$ is the standard deviation of the light curve, and $\sigma_{\rm t}$ is calculated from the standard deviation of the residuals between the observational data and the smoothed light curves. The error of $m_{\text{var}}$ is given by
\begin{equation}
\sigma_{m_{\rm var}} = \frac{S^2}{m_{\rm var}}\sqrt{\frac{1}{2N}},
\end{equation}
where $N$ represents the number of data points in each light curve. 

Sources with less than three data points in the MIR or optical bands between two spectroscopic observations are excluded from the analysis. Consequently, 72 targets are retained. Among these, the number of targets exhibiting variability amplitudes greater than 0.1 mag in each band is as follows: W1 (41/72), W2 (48/72), and optical (62/72). The mean intrinsic variability amplitudes in the W1, W2, and optical bands are 0.12, 0.15, and 0.26 mag, respectively. The variability amplitude of W1, W2, and optical bands are presented in Table \ref{tab:example} (in Appendix \ref{Appendix_table}).

\subsubsection{Variability in the optical band}
In the optical band, 20 sources display clear variability structures upon visual inspection, with three objects showing maximum magnitude variations exceeding $1\, \mathrm{mag}$ (J095536.70+103752.4, J161711.42+063833.5, and J151143.41+210104.0). However, more than 60 sources do not exhibit discernible long-term variability in their optical light curves. This lack of variability could be attributed to several factors:

\begin{itemize}
    \item Sparse Data Sampling: For {about 7} sources, half of time between two spectroscopic observations are not covered by photometric observations. The limited number of optical data points makes it difficult to identify variability structures.
    \item High Photometric Scatter: In certain cases, the photometric data exhibit dispersions exceeding $0.3\, \mathrm{mag}$, which dominate the light curves and obscure the underlying long-term variability.
    \item Host Galaxy Contribution: As our sample primarily consists of low-redshift AGNs, the optical continuum often includes a significant contribution from the host galaxy. Spectral fitting indicates that, for most sources, the host galaxy accounts for over 50\% of the optical flux, significantly diluting the observed variability amplitude.
    \item Intrinsic Properties: A small number of objects may inherently lack significant variability in the optical band during the photometric observations.
\end{itemize}

\subsubsection{Variability in the MIR band}
The MIR emission observed in AGNs is generally attributed to the reprocessing of radiation from the central source's energy by the surrounding torus \citep{2013peag.book.....N}. As stellar emission in the MIR is relatively negligible, compared to optical wavelengths, MIR variability more directly reflects changes in the luminosity of the central source and is less affected by other components of the host galaxy. As previously mentioned, we have collected WISE light curves for our targets. Compared to the optical band, the WISE data reveal more pronounced variability features. Among the 20 targets showing clear optical variability, all also exhibit significant variability in the WISE bands. Additionally, among the remaining targets, approximately 50 exhibit significant MIR variability, with 22 of these showing maximum magnitude variations exceeding 0.5 magnitude. Notably, two sources (J120447.92+170256.9, and J151143.41+210104.0) display maximum magnitude variations exceeding 1 magnitude. The targets that do not show significant variability in either the optical or MIR bands may have not captured the variability events. Our findings suggest that for low-redshift AGNs, the MIR is more suitable for revealing the variability characteristics of changing-look quasars, aligning with the results reported by \citet{2017ApJ...846L...7S}.
 
We exclude sources with less than 3 valid data points in the W1 and W2 bands between two spectroscopic observations. The variability amplitudes in the W1 and W2 bands of 72 remaining sources are calculated according to Equation (7). Figure \ref{w1-w2} shows that the variability amplitudes in the W2 band of 56 sources are greater than those in the W1 band, exhibiting a `redder when brighter' trend, which is consistent with \citet{yang2018discovery}. The radiation of the W1 and W2 bands originate from the re-radiation of the heated dust torus, and the emission region of the W2 band is farther from the central engine than the emission region of the W1 band \citep{2019ApJ...886..150M,2023MNRAS.522.3439C,2024ApJ...964...37L,2024ApJ...968...59M}. The size of the dust torus is determined by the luminosity of the central source \citep[e.g.,][]{1987ApJ...320..537B}. A plausible explanation is that the increased radiation from the central source heats the dust clouds, causing them to sublimate and thereby expanding the inner boundary of the torus. We consider that with the same radial expansion, the emission region of the W2 bands increases more significantly than the emission region of the W1 bands. Thus, the emission region of the W2 band has a more significant change, resulting in a greater variability amplitude than W1.

\subsection{Repeating CL~AGN and candidates}

\begin{figure*}
  \begin{center}
    \resizebox{110mm}{!} {\includegraphics *{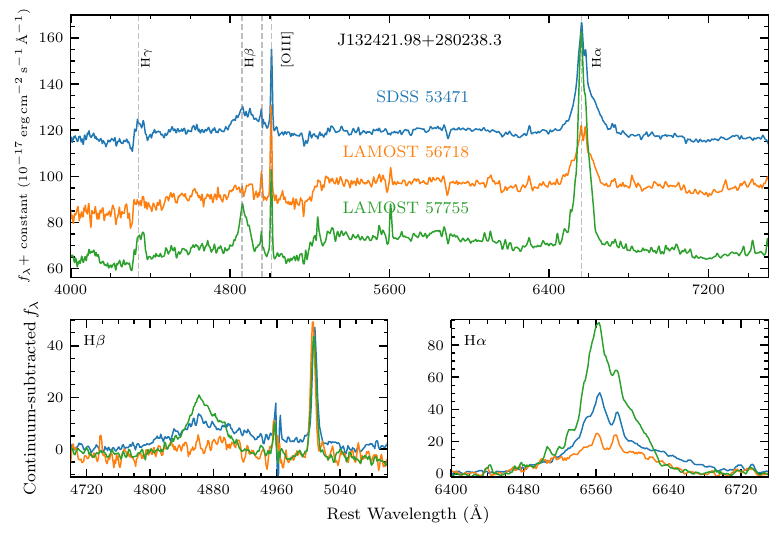}}
    \resizebox{60mm}{!} {\includegraphics *{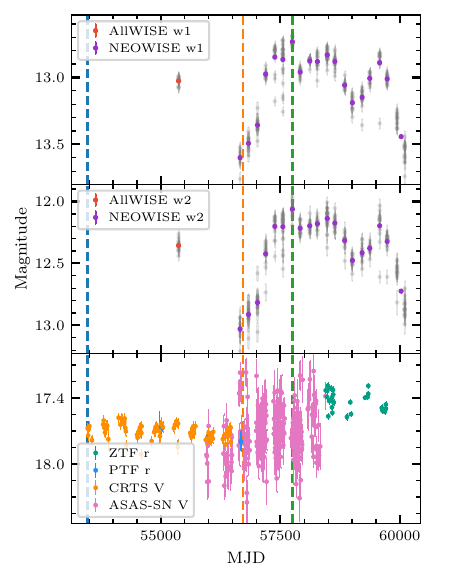}}
\caption{The spectra and light curves of RCL AGN, J132421.98+280238.3. The left panel shows the multi-epoch spectra, \hb, and \ha\ emission line profiles. The right panel shows the light curves and the colored dotted lines correspond to three spectral observations.}
\label{fig_Repeat_CL_AGN}
\end{center}
\end{figure*}

Figure \ref{fig_Repeat_CL_AGN} presents the multi-epoch spectroscopic observations of J132421.98+280238.3. The LAMOST spectra are flux-calibrated using the SDSS spectrum as a reference. The initial SDSS spectrum obtained at MJD 53471 exhibits prominent broad \hb\ and \ha\ emission lines. The subsequent spectrum taken at MJD 56718 exhibits a dramatic spectral transition: the \hb\ emission line disappears and the \ha\ emission line become weaker, indicating a turn-off behavior. The later LAMOST observation at MJD 57755 displays a full recovery of the \hb\ emission line, with even greater strength than in the initial epoch, confirming a turn-on behavior between MJD 56718 and 57755. Based on these spectral transitions, we identify J132421.98+280238.3 as a RCL AGN.

As shown in Figure \ref{fig_Repeat_CL_AGN}, the disappearance and reappearance of \hb\ emission line between MJD 53471 and MJD 57755 is consistent with the dimming and rebrightening of MIR light curves. Additionally, the MIR light curves exhibit a decline after MJD 57755, potentially indicating another turn-off behavior. These repeated transitions are likely driven by changes of accretion rate rather than dust obscuration \citep{2025ApJ...981..129W}. 

\begin{deluxetable}{lccccB}
\tablewidth{0pt}
\tablecaption{RCL~AGN candidates \label{table_RCL_AGNs}}
\centering
\tablehead{\colhead{Name} & \colhead{Transition} & \colhead{W1} & \colhead{W2} & \colhead{r-band}
}
\startdata
J021359.78+004226.8 & Turn-off & 0.77 & 0.77 & 17.63 \\ 
J084016.42+220624.6 & Turn-off & 0.37 & 0.56 & 18.79 \\ 
J094838.42+403043.7 & Turn-off & 0.34 & 0.38 & 16.23 \\ 
J104043.38+273518.9 & Turn-off & 0.44 & 0.44 & 17.97 \\ 
J104705.16+544405.9 & Turn-off & 0.44 & 0.51 & 19.01 \\ 
J105233.83+454949.3 & Turn-off & 0.25 & 0.27 & 17.95 \\ 
J113402.78+290403.9 & Turn-off & 0.54 & 0.53 & 18.68 \\ 
J114634.92+282642.0 & Turn-off & 0.20 & 0.23 & 18.36 \\ 
J122623.69+261050.0 & Turn-off & 0.42 & 0.56 & 17.11 \\ 
J123359.13+084211.6 & Turn-off & 0.29 & 0.35 & 18.27 \\ 
J132421.98+280238.3 & Turn-off & 0.74 & 0.76 & 16.93 \\ 
J134241.02+321231.5 & Turn-off & 0.30 & 0.51 & 18.43 \\ 
J134739.65+362240.0 & Turn-off & 0.60 & 0.68 & 18.78 \\ 
J161711.42+063833.5 & Turn-off & 0.51 & 0.43 & 18.86 \\ 
J172533.07+571645.6 & Turn-off & 0.27 & 0.38 & 17.09 \\ 
J083132.25+364617.3 & Turn-on  & 0.38 & 0.66 & 18.42 \\ 
J085359.05+212803.4 & Turn-on  & 0.89 & 0.98 & 16.92 \\ 
J093730.32+260232.1 & Turn-on  & 0.44 & 0.69 & 18.08 \\ 
\enddata
\tablenotetext{  }{NOTE-Columns: (1) Object name. (2) Transition type. (3) The maximum magnitude variations of the W1 band after the second spectroscopic observation. (4) The maximum magnitude variations of the W2 band after the second spectroscopic observation. (5) The apparent magnitude of r-band.}
\end{deluxetable}

\begin{figure*}
  \begin{center}
    \resizebox{75mm}{!} {\includegraphics *{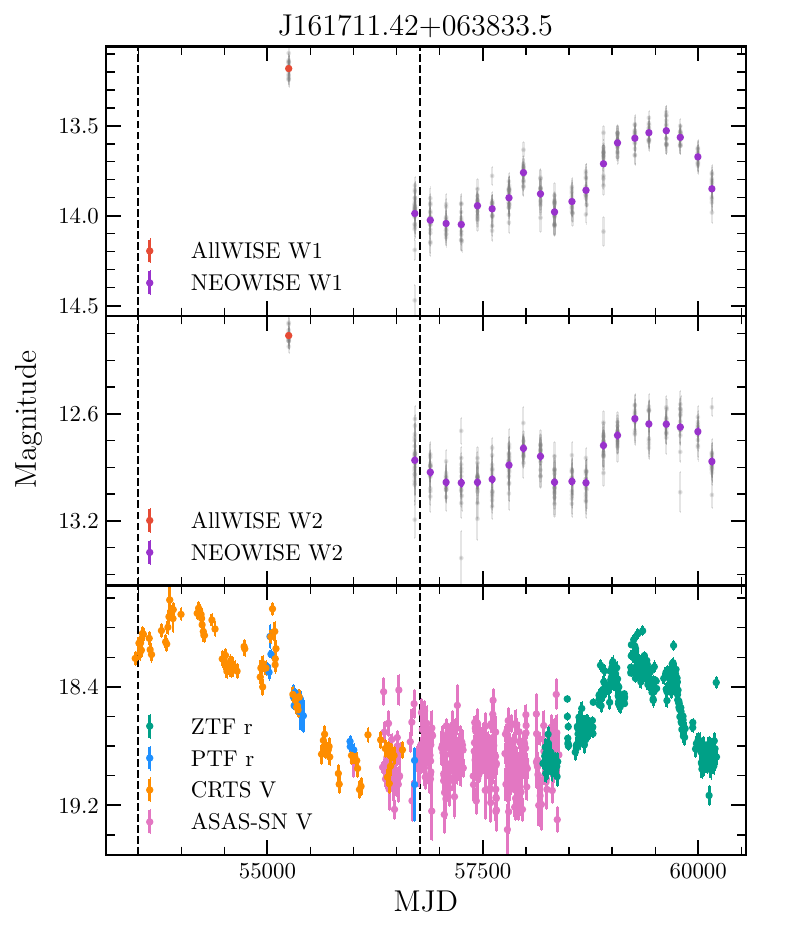}}
    \resizebox{75mm}{!} {\includegraphics *{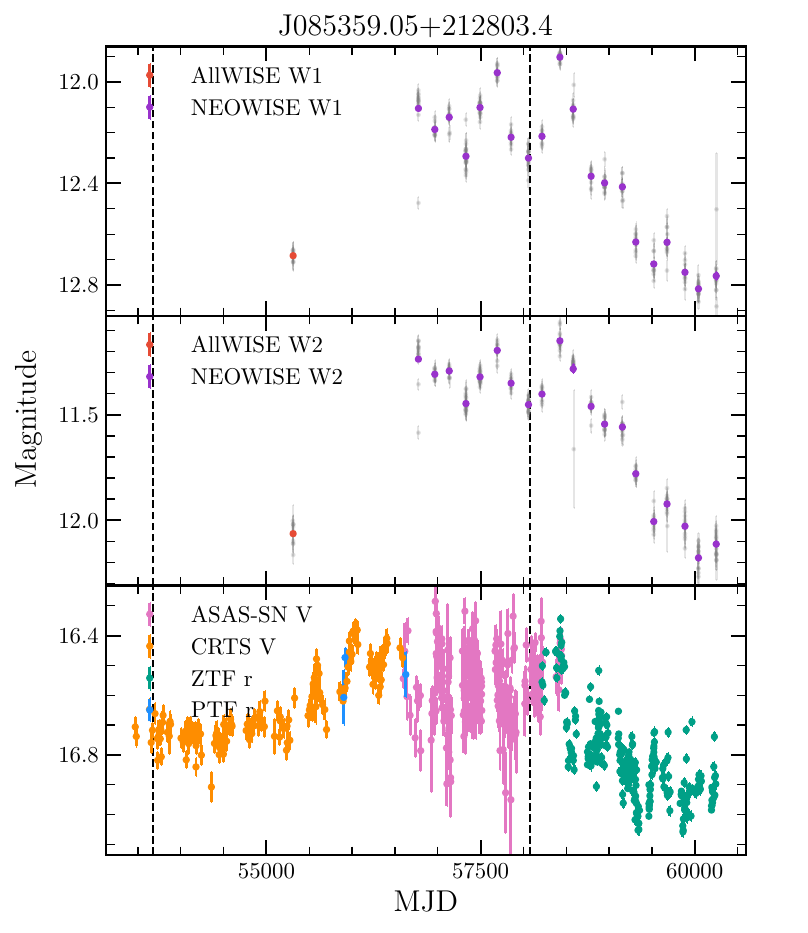}}
    \caption {The light curves in MIR and optical band for J161711.42+063833.5 (left panel), J085359.05+212803.4 (right panel). The vertical dashed lines represent the observation time of two optical spectra. The turn-off CL~AGN J161711.42+063833.5 becomes brighter after the CL phenomenon. The turn-on CL~AGN J085359.05+212803.4 becomes dimmer after the second optical spectrum.}
\label{Lihtcurve_spectrial}
\end{center}
\end{figure*}

Additionally, we find that the recent WISE light curves of some targets exhibit variability patterns, that are opposite to those observed during their initial CL phase. Turn-on (turn-off) CL~AGNs exhibit significantly decreasing (increasing) behaviors in recent WISE fluxes. We name this type of behavior as inverted variability. We find 18 sources exhibit inverted variability after their initial CL phenomenon. For typical AGNs, the optical continuum variability amplitude is about $0.2\,\rm{mag}$ \citep{2004ApJ...601..692V,2007AJ....134.2236S,2018ApJ...854..160R}. The maximum magnitude variations of the inverted variability exceed $0.2\,\rm{mag}$ and exhibit long-term trends. These are highlighted in Table \ref{table_RCL_AGNs} and considered as repeating CL~AGN (RCL~AGN) candidates. Fifteen candidates exhibit significantly increasing behavior in the MIR band following the second spectral observation (e.g., J161711.42+063833.5, whose light curves are shown in the left panel of Figure \ref{Lihtcurve_spectrial}.), suggesting the potential for a turn-on phenomenon. Three candidates exhibit decreasing behavior (e.g., J085359.05+212803.4, whose light curves are shown in the right panel of Figure \ref{Lihtcurve_spectrial}.). These are promising RCL~AGN candidates that can be verified by future spectroscopic observations.

Among RCL~AGN candidates, J161711.42+063833.5 was first identified as CL~AGN by \cite{2019ApJ...874....8M} and later confirmed as an RCL~AGN by \cite{wang2024dormancyreawakeningyearsnew}. \cite{wang2024dormancyreawakeningyearsnew} reported the presence of the \hb\ emission line in the follow-up spectra obtained on June 7, 2022 (MJD = 59738). MIR light curves show a decreasing trend after the spectral observation as shown in the left panel of Figure \ref{Lihtcurve_spectrial}, possibly signaling another turn-off event. Similar patterns are also observed in J122623.69+261050.0 and J132421.98+280238.3. To identify the RCL~AGN candidates, we are requesting follow-up spectroscopic observations of them.

\section{Summary}\label{sec:summary}
In this study, we have identified CL~AGNs by cross-matching repeat observations of AGN and galaxy spectra from the SDSS DR18 and LAMOST surveys. We have focused on the CL phenomenon in the \hb\ emission line, selecting samples with a redshift of less than 0.8. First, we fitted the spectra of repeated observed AGNs and galaxies, using the fitting parameters to identify potential CL~AGN sources. These sources were then manually inspected, resulting in a final sample of CL~AGNs. The main results of our study are as follows:
\begin{enumerate}
    \item We have identified a sample of 51 CL~AGNs. Among them, 40 are newly discovered CL~AGNs. Of the 51 CL~AGNs, 44 are turn-off type, and 7 are turn-on type. Additionally, we have identified 31 candidates, two of which are identified as CL AGN in previous studies.
    
    \item Within our sample, 41 CL~AGNs exhibited CL behaviors only in the \hb\ emission line, 9 showed changes in both \hb\ and \ha\ emission lines, and one source displayed CL characteristics only in the \ha\ emission line. For objects that only exhibited changes in the \hb\ emission line, their \ha\ BEL was consistently present. Conversely, for AGNs that only exhibited changes in the \ha\ emission line, the \hb\ emission line was always observed as a narrow line. This result is consistent with the findings of \citet{guohx2019}, which align with the emission line sequence characteristics of CL~AGNs.
    
    \item We have estimated the black hole masses and Eddington ratios of the sample. The black hole mass range from $2.54\times 10^6 \, M_{\odot}$ to $8.03\times10^8 \, M_\odot$. Compared to large samples of AGNs \citep{Shen_2011}, our sample exhibited lower Eddington ratios ($L_\mathrm{bol}/L_\mathrm{Edd}$ values ranging from 0.001 to 0.13), consistent with previous studies. The result indicates that CL~AGNs generally have lower accretion rates, and suggests that the CL phenomenon is likely due to state transitions in the accretion disk.
    
    \item The variability amplitude in the W2 band exceeds that of W1, aligning with the findings of \citet{yang2018discovery}. We suggest that this phenomenon may arise from more significant changes in the emission region of the W2 band in response to variation amplitude.

    \item We identified J132421.98+280238.3 as a RCL AGN, based on the multi-epoch spectra obtained from SDSS and LAMOST.
    
    \item We selected 18 RCL~AGN candidates from our sample, which are turn-off (turn-on) CL~AGNs with recent WISE fluxes significantly increasing (decreasing). Among these, J161711.42+063833.5 previously identified by \citet{2019ApJ...874....8M}, has been confirmed as an RCL~AGN by \citet{wang2024dormancyreawakeningyearsnew}.
\end{enumerate}

\section{Acknowledgments}
We thank Grisha Zeltyn for providing the CL~AGN catalog of their study \citep{2024ApJ...966...85Z}, which enables us to conduct a comprehensive check for duplicate sources and newly discovered sources. We also thank Xue-Bing Wu for helpful discussion, and thank the anonymous referee for constructive comments that improved the paper. This work was supported by the National Natural Science Foundation of China under grants 12033006, 11925301, 12433007, 12221003, 12263003, 12322303, and 12363002. 

We acknowledge the data provided by LAMOST. Guoshoujing Telescope (the Large Sky Area Multi-Object Fiber Spectroscopic Telescope LAMOST) is a National Major Scientific Project built by the Chinese Academy of Sciences. Funding for the project has been provided by the National Development and Reform Commission. LAMOST is operated and managed by the National Astronomical Observatories, Chinese Academy of Sciences.

We acknowledge the data provided by SDSS. SDSS is managed by the Astrophysical Research Consortium for the Participating Institutions of the SDSS Collaboration including the Brazilian Participation Group, the Carnegie Institution for Science, Carnegie Mellon University, Center for Astrophysics Harvard \& Smithsonian, the Chilean Participation Group, the French Participation Group, Instituto de Astrofísica de Canarias, The Johns Hopkins University, Kavli Institute for the Physics and Mathematics of the Universe (IPMU) / University of Tokyo, the Korean Participation Group, Lawrence Berkeley National Laboratory, Leibniz Institut für Astrophysik Potsdam (AIP), Max-Planck-Institut für Astronomie (MPIA Heidelberg), Max-Planck-Institut für Astrophysik (MPA Garching), Max-Planck-Institut für Extraterrestrische Physik (MPE), National Astronomical Observatories of China, New Mexico State University, New York University, University of Notre Dame, Observatório Nacional / MCTI, The Ohio State University, Pennsylvania State University, Shanghai Astronomical Observatory, United Kingdom Participation Group, Universidad Nacional Autónoma de México, University of Arizona, University of Colorado Boulder, University of Oxford, University of Portsmouth, University of Utah, University of Virginia, University of Washington, University of Wisconsin, Vanderbilt University, and Yale University.

This paper also contains the data collected by WISE, CRTS, PTF, ZTF, and ASAS-SN. The CRTS survey is supported by the U.S. National Science Foundation under grants AST-0909182 and AST-1313422. ZTF is Supported by the National Science Foundation under Grants No. AST-1440341 and AST-2034437 and a collaboration including current partners Caltech, IPAC, the Oskar Klein Center at Stockholm University, the University of Maryland, University of California, Berkeley, the University of Wisconsin at Milwaukee, University of Warwick, Ruhr University, Cornell University, Northwestern University and Drexel University. Operations are conducted by COO, IPAC, and UW. 

This publication also uses datas from WISE, which is a joint project of the University of California, Los Angeles, and the Jet Propulsion Laboratory/California Institute of Technology, funded by the National Aeronautics and Space Administration. This publication also makes use of data products from NEOWISE, which is a project of the Jet Propulsion Laboratory/California Institute of Technology, funded by the Planetary Science Division of the National Aeronautics and Space Administration.

\software{Astropy \citep{2013A&A...558A..33A,2018AJ....156..123A,2022ApJ...935..167A}, QSOFITMORE \citep{fu2021finding}, TOPCAT \citep{2005ASPC..347...29T}.}

\appendix

\section{Low SNR spectra}
\setcounter{figure}{0}
\renewcommand{\thefigure}{A\arabic{figure}}
\label{Appendix_low_SNR}

We present the spectral fitting results of CL~AGN, J114634.92+282642.0. The median SNRs of SDSS spectrum and LAMOST spectra are 14.5 and 6.3, respectively. Significant changes are observed in the \hb\ and \ha\ emission lines between the two epochs, exhibiting a clear turn-off behavior.

\begin{figure*}
  \begin{center}
    \resizebox{89mm}{!} {\includegraphics *{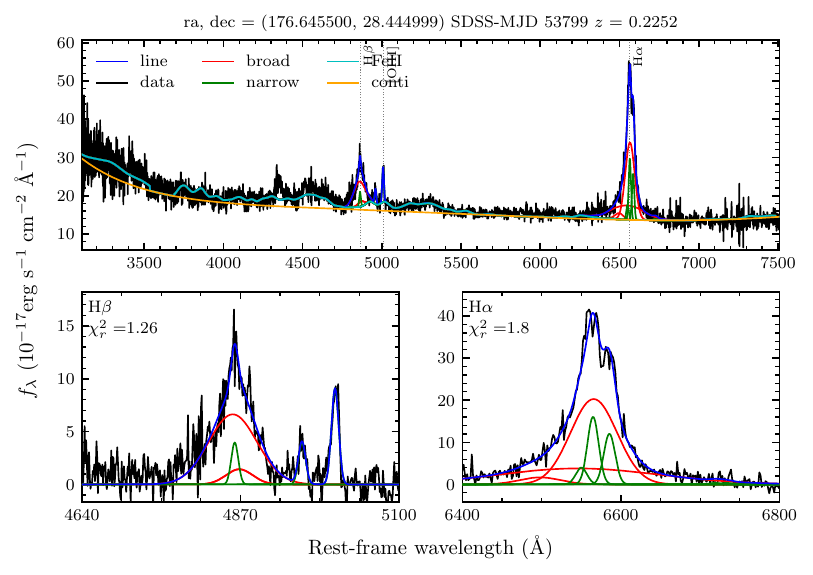}}
    \resizebox{89mm}{!} {\includegraphics *{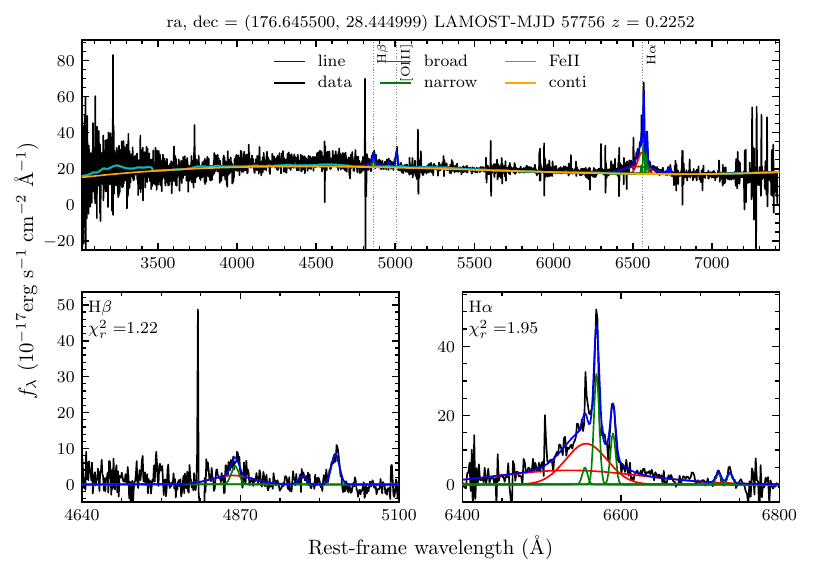}}
\caption{The spectral fitting results of the SDSS and LAMOST spectra for the CL AGN, J114634.92+282642.0. }
\label{fig:low_SNR}
\end{center}
\end{figure*}

\section{The basic properties of CL AGNs and candidates}
\label{Appendix_table}
\setcounter{table}{0}
\renewcommand{\thetable}{B\arabic{table}}
We present the basic properties of 51 CL AGNs and 31 candidates. 

\begin{longrotatetable}
\setlength{\tabcolsep}{1pt}
\tabletypesize{\scriptsize}  
\begin{deluxetable}{cccccccccccccccccc}
\tablecaption{ The basic properties of the CL~AGNs and candidates in our samples \label{tab:example}}
\tablehead{\colhead{Name} & \colhead{R.A.} & \colhead{Dec.}& \colhead{z}& \colhead{ $\mathrm{MJD_1}$ }& \colhead{ $\mathrm{MJD_2}$ } & \colhead{$\log M_{\mathrm{BH}}$} & \colhead{$\log L_{\mathrm{bol,1}}$}& \colhead{$\log L_{\mathrm{bol,2}}$}& \colhead{ $\log\lambda_\mathrm{Edd,1}$} & \colhead{ $\log\lambda_\mathrm{Edd,2}$ } & $m_{\mathrm{var,r}}$ &  $m_{\mathrm{var,w1}}$ & $m_{\mathrm{var,w2}}$ & \colhead{Line} & \colhead{Type}  & \colhead{Note} & \colhead{Ref}  \\
 \colhead{} &  \colhead{} &  \colhead{} &  \colhead{} &  \colhead{} &  \colhead{} &  \colhead{$M_{\odot}$} &   \colhead{$\mathrm{erg\,s^{-1}}$} &  \colhead{$\mathrm{erg\,s^{-1}}$} &  \colhead{} &  \colhead{} &  \colhead{}&  \colhead{}&  \colhead{}&  \colhead{}&  \colhead{}&  \colhead{} &  \colhead{}}
\startdata
J000253.52+210109.9	&	0.723	&	21.0194	&	0.1464	&	56957	&	57360	&	7.49$\pm$0.01	&	44.44$\pm$0.01	&	44.1$\pm$0.03	&	-1.15$\pm$0.01	&	-1.49$\pm$0.03	&	…	&	…	&	…	&	H$\beta$	&	Turn-off	&   CL AGN &		\\
J004658.26+092045.7	&	11.7427	&	9.346	&	0.1606	&	58424	&	58465	&	7.44$\pm$0.01	&	44.38$\pm$0.01	&	44.32$\pm$0.02	&	-1.16$\pm$0.01	&	-1.23$\pm$0.02	&	…	&	…	&	…	&	H$\beta$	&	Turn-off	&   Candidate &		\\
J021359.78+004226.8	&	33.4991	&	0.7074	&	0.1824	&	51816	&	57747	&	8.35$\pm$0.01	&	45.36$\pm$0	&	45.05$\pm$0.06	&	-1.09$\pm$0.01	&	-1.4$\pm$0.06	&	0.07 $\pm$ 0.03	&	0.11 $\pm$ 0.03	&	0.07 $\pm$ 0.02	&	H$\beta$	&	Turn-off & CL AGN	&	(1)	\\
J083132.25+364617.3	&	127.8844	&	36.7715	&	0.195	&	52312	&	57367	&	8.26$\pm$0.09	&	44.12$\pm$0.02	&	44.62$\pm$0.01	&	-2.24$\pm$0.09	&	-1.74$\pm$0.09	&	0.24 $\pm$ 0.06	&	0.17 $\pm$ 0.06	&	0.24 $\pm$ 0.09	&	H$\beta$	&	Turn-on  & CL AGN	&	(2)	\\
J083826.50+371906.7	&	129.6104	&	37.3185	&	0.2111	&	52320	&	59200	&	8.72$\pm$0.04	&	44.82$\pm$0.01	&	45.05$\pm$0.01	&	-2$\pm$0.04	&	-1.77$\pm$0.04	&	0.14 $\pm$ 0.04	&	0.15 $\pm$ 0.04	&	0.14 $\pm$ 0.03	&	H$\beta$	&	Turn-on	  & CL AGN &		\\
J084016.42+220624.6	&	130.0684	&	22.1068	&	0.2044	&	53360	&	58136	&	7.66$\pm$0.05	&	44.63$\pm$0.02	&	44.24$\pm$0.03	&	-1.12$\pm$0.05	&	-1.51$\pm$0.05	&	0.24 $\pm$ 0.04	&	0.18 $\pm$ 0.04	&	0.24 $\pm$ 0.06	&	H$\beta$	&	Turn-off  & CL AGN	&		\\
J084635.23+465039.2	&	131.6468	&	46.8442	&	0.1468	&	52235	&	57459	&	7.98$\pm$0.02	&	44.64$\pm$0.01	&	44.03$\pm$0.03	&	-1.44$\pm$0.02	&	-2.05$\pm$0.03	&	0.14 $\pm$ 0.05	&	0.14 $\pm$ 0.05	&	0.14 $\pm$ 0.05	&	$\rm{H\beta}$, $\rm{H\alpha}$	&	Turn-off	&   Candidate  &		\\
J084957.78+274729.0	&	132.4908	&	27.7914	&	0.2985	&	53350	&	56628	&	7.82$\pm$0.1	&	45$\pm$0.04	&	44.32$\pm$0.07	&	-0.92$\pm$0.1	&	-1.59$\pm$0.11	&	…	&	…	&	…	&	H$\beta$	&	Turn-off  & CL AGN &	(2)	\\
J085359.05+212803.4	&	133.4961	&	21.4676	&	0.0842	&	53680	&	58078	&	7.83$\pm$0.03	&	44.13$\pm$0.01	&	44.37$\pm$0.01	&	-1.8$\pm$0.03	&	-1.56$\pm$0.03	&	0.25 $\pm$ 0.05	&	0.18 $\pm$ 0.05	&	0.25 $\pm$ 0.06	&	H$\beta$	&	Turn-on	  & CL AGN &		\\
J090204.29+182605.1	&	135.5179	&	18.4348	&	0.0922	&	53729	&	58106	&	7.58$\pm$0.03	&	44.01$\pm$0.01	&	43.62$\pm$0.07	&	-1.67$\pm$0.03	&	-2.05$\pm$0.08	&	0.09 $\pm$ 0.03	&	0.1 $\pm$ 0.03	&	0.09 $\pm$ 0.04	&	H$\beta$	&	Turn-off  &   Candidate 	&		\\
J090455.96+300613.3	&	136.2332	&	30.1037	&	0.1086	&	52974	&	57724	&	7.34$\pm$0.01	&	44.22$\pm$0.03	&	42.54$\pm$0.17	&	-1.23$\pm$0.03	&	-2.91$\pm$0.17	&	0.18 $\pm$ 0.02	&	0.05 $\pm$ 0.02	&	0.18 $\pm$ 0.05	&	$\rm{H\beta}$, $\rm{H\alpha}$	&	Turn-off	&   Candidate  &		\\
J091730.71+110110.7	&	139.378	&	11.0196	&	0.3036	&	53050	&	58518	&	8.57$\pm$0.04	&	45.15$\pm$0.02	&	44.92$\pm$0.02	&	-1.52$\pm$0.04	&	-1.75$\pm$0.05	&	0.11 $\pm$ 0.02	&	0.11 $\pm$ 0.02	&	0.11 $\pm$ 0.03	&	H$\beta$	&	Turn-off &   Candidate	&		\\
J093005.31+173925.6	&	142.5221	&	17.6571	&	0.2164	&	53728	&	59551	&	8.25$\pm$0.02	&	44.9$\pm$0.01	&	44.78$\pm$0.01	&	-1.46$\pm$0.02	&	-1.58$\pm$0.03	&	0.2 $\pm$ 0.04	&	0.16 $\pm$ 0.04	&	0.2 $\pm$ 0.04	&	H$\beta$	&	Turn-off  &   Candidate &		\\
J093327.87+093200.1	&	143.3661	&	9.5334	&	0.0498	&	52993	&	56394	&	7.29$\pm$0.02	&	43.54$\pm$0.01	&	42.49$\pm$0.09	&	-1.86$\pm$0.02	&	-2.9$\pm$0.09	&	…	&	…	&	…	&	$\rm{H\beta}$, $\rm{H\alpha}$	&	Turn-off	&   Candidate &		\\
J093730.32+260232.1	&	144.3764	&	26.0423	&	0.1623	&	53733	&	57369	&	7.58$\pm$0.02	&	44$\pm$0.01	&	44.49$\pm$0.03	&	-1.68$\pm$0.03	&	-1.19$\pm$0.04	&	0.12 $\pm$ 0.04	&	0.09 $\pm$ 0.04	&	0.12 $\pm$ 0.05	&	H$\beta$	&	Turn-on	 & CL AGN&	(2)	\\
J094144.82+575123.7	&	145.4368	&	57.8566	&	0.1586	&	51911	&	57039	&	8.29$\pm$0.05	&	44.63$\pm$0.01	&	44.03$\pm$0.03	&	-1.76$\pm$0.05	&	-2.36$\pm$0.05	&	0.26 $\pm$ 0.08	&	0.18 $\pm$ 0.08	&	0.26 $\pm$ 0.11	&	H$\alpha$	&	Turn-off &   Candidate	&		\\
J094838.42+403043.7	&	147.1601	&	40.5121	&	0.0468	&	52709	&	58461	&	7.41$\pm$0.04	&	44.12$\pm$0.01	&	44.03$\pm$0.01	&	-1.39$\pm$0.04	&	-1.48$\pm$0.04	&	0.09 $\pm$ 0.01	&	0.05 $\pm$ 0.01	&	0.09 $\pm$ 0.02	&	H$\beta$	&	Turn-off	& CL AGN &	(3)	\\
J095536.70+103752.4	&	148.9029	&	10.6312	&	0.2844	&	52996	&	59200	&	8.39$\pm$0.04	&	45.26$\pm$0.01	&	45.56$\pm$0.02	&	-1.23$\pm$0.04	&	-0.93$\pm$0.04	&	0.15 $\pm$ 0.04	&	0.22 $\pm$ 0.04	&	0.15 $\pm$ 0.03	&	H$\beta$	&	Turn-on	 & CL AGN &		\\
J101403.44+193609.1	&	153.5143	&	19.6025	&	0.1149	&	53765	&	57477	&	7.29$\pm$0.11	&	44.09$\pm$0.05	&	43.63$\pm$0.02	&	-1.3$\pm$0.12	&	-1.75$\pm$0.11	&	0.14 $\pm$ 0.04	&	0.12 $\pm$ 0.04	&	0.14 $\pm$ 0.05	&	H$\beta$	&	Turn-off & CL AGN &		\\
J101417.87+351106.8	&	153.5745	&	35.1852	&	0.4151	&	53357	&	58897	&	…	&	…	&	…	&	…	&	…	&	0.26 $\pm$ 0.04	&	0.17 $\pm$ 0.04	&	0.26 $\pm$ 0.06	&	H$\beta$	&	Turn-off & CL AGN	&		\\
J101628.00+463051.6	&	154.1167	&	46.5143	&	0.1408	&	52614	&	59550	&	7.82$\pm$0.1	&	44.52$\pm$0.06	&	44.71$\pm$0.05	&	-1.4$\pm$0.11	&	-1.21$\pm$0.11	&	0.09 $\pm$ 0.02	&	0.07 $\pm$ 0.02	&	0.09 $\pm$ 0.02	&	$\rm{H\beta}$, $\rm{H\alpha}$	&	Turn-off &   Candidate	&		\\
J102822.85+235125.8	&	157.0952	&	23.8572	&	0.1736	&	53734	&	57103	&	8.01$\pm$0.02	&	44.65$\pm$0	&	44.17$\pm$0.02	&	-1.46$\pm$0.02	&	-1.95$\pm$0.02	&	0.12 $\pm$ 0.07	&	0.07 $\pm$ 0.07	&	0.12 $\pm$ 0.11	&	$\rm{H\beta}$, $\rm{H\alpha}$	&	Turn-off	& CL AGN&		\\
J103359.13+521047.7	&	158.4964	&	52.1799	&	0.1043	&	52644	&	58872	&	7.18$\pm$0.01	&	43.87$\pm$0.01	&	43.52$\pm$0.1	&	-1.41$\pm$0.02	&	-1.77$\pm$0.1	&	0.16 $\pm$ 0.03	&	0.11 $\pm$ 0.03	&	0.16 $\pm$ 0.04	&	$\rm{H\beta}$, $\rm{H\alpha}$	&	Turn-off	&   Candidate &		\\
J103844.05+390804.0	&	159.6835	&	39.1344	&	0.1165	&	53003	&	58926	&	8.19$\pm$0.03	&	44.46$\pm$0.01	&	44.24$\pm$0.03	&	-1.84$\pm$0.03	&	-2.06$\pm$0.04	&	0.06 $\pm$ 0.03	&	0.06 $\pm$ 0.03	&	0.06 $\pm$ 0.03	&	H$\beta$	&	Turn-off &CL AGN	&		\\
J104043.38+273518.9	&	160.1808	&	27.5886	&	0.1671	&	53797	&	56685	&	7.89$\pm$0.02	&	44.66$\pm$0	&	44.46$\pm$0.01	&	-1.33$\pm$0.02	&	-1.53$\pm$0.02	&	…	&	…	&	…	&	H$\beta$	&	Turn-off  & CL AGN	&		\\
J104111.56+555142.4	&	160.2982	&	55.8618	&	0.3225	&	52368	&	57870	&	8.11$\pm$0.05	&	45.07$\pm$0.02	&	44.84$\pm$0.04	&	-1.13$\pm$0.06	&	-1.36$\pm$0.07	&	0.09 $\pm$ 0.03	&	0.11 $\pm$ 0.03	&	0.09 $\pm$ 0.03	&	H$\beta$	&	Turn-off  & CL AGN	&		\\
J104332.88+010108.8	&	160.887	&	1.0191	&	0.0719	&	51910	&	57394	&	7.3$\pm$0.01	&	44.14$\pm$0.01	&	43.59$\pm$0.02	&	-1.27$\pm$0.01	&	-1.82$\pm$0.02	&	0.23 $\pm$ 0.06	&	0.17 $\pm$ 0.06	&	0.23 $\pm$ 0.08	&	$\rm{H\beta}$, $\rm{H\alpha}$	&	Turn-off &   Candidate	&		\\
J104427.75+271806.3	&	161.1156	&	27.3018	&	0.0758	&	53786	&	58133	&	7.74$\pm$0.06	&	44.44$\pm$0.02	&	43.94$\pm$0	&	-1.4$\pm$0.06	&	-1.91$\pm$0.06	&	0.09 $\pm$ 0.02	&	0.08 $\pm$ 0.02	&	0.09 $\pm$ 0.02	&	$\rm{H\beta}$, $\rm{H\alpha}$	&	Turn-off  &CL AGN	&		\\
J104705.16+544405.9	&	161.7715	&	54.735	&	0.2149	&	52368	&	57044	&	8.11$\pm$0.07	&	45.28$\pm$0.02	&	45.3$\pm$0.01	&	-0.92$\pm$0.07	&	-0.9$\pm$0.07	&	0.13 $\pm$ 0.05	&	0.12 $\pm$ 0.05	&	0.13 $\pm$ 0.05	&	H$\beta$	&	Turn-off & CL AGN	&	(4)	\\
J104913.80+044040.0	&	162.3075	&	4.6778	&	0.0885	&	52338	&	57394	&	7.34$\pm$0.02	&	43.95$\pm$0.01	&	...	&	-1.49$\pm$0.02	&	...	&	0.05 $\pm$ 0.02	&	0.04 $\pm$ 0.02	&	0.05 $\pm$ 0.11	&	$\rm{H\beta}$, $\rm{H\alpha}$	&	Turn-off&   Candidate	&		\\
J105233.83+454949.3	&	163.1409	&	45.8304	&	0.1267	&	53053	&	58171	&	7.58$\pm$0.07	&	44.11$\pm$0.01	&	43.84$\pm$0.01	&	-1.57$\pm$0.07	&	-1.85$\pm$0.07	&	0.14 $\pm$ 0.02	&	0.07 $\pm$ 0.02	&	0.14 $\pm$ 0.04	&	H$\beta$	&	Turn-off & CL AGN	&		\\
J110413.32+281530.3	&	166.0555	&	28.2584	&	0.1866	&	53786	&	57377	&	7.88$\pm$0.12	&	44.9$\pm$0.04	&	44.3$\pm$0.02	&	-1.08$\pm$0.13	&	-1.68$\pm$0.12	&	0.25 $\pm$ 0.07	&	0.21 $\pm$ 0.07	&	0.25 $\pm$ 0.08	&	H$\beta$	&	Turn-off  & CL AGN	&		\\
J112229.65+214815.6	&	170.6236	&	21.8043	&	0.0602	&	54178	&	57453	&	7.07$\pm$0.02	&	43.67$\pm$0.02	&	43.48$\pm$0.02	&	-1.5$\pm$0.03	&	-1.69$\pm$0.03	&	0.1 $\pm$ 0.03	&	0.06 $\pm$ 0.03	&	0.1 $\pm$ 0.04	&	H$\beta$	&	Turn-off  & CL AGN		&		\\
J113229.14+035729.1	&	173.1214	&	3.9581	&	0.0909	&	52642	&	57392	&	7.45$\pm$0.07	&	43.23$\pm$0.03	&	43.79$\pm$0.03	&	-2.32$\pm$0.07	&	-1.76$\pm$0.07	&	0.25 $\pm$ 0.06	&	0.18 $\pm$ 0.06	&	0.25 $\pm$ 0.08	&	H$\beta$	&	Turn-on  & CL AGN		&	(2)	\\
J113402.78+290403.9	&	173.5116	&	29.0678	&	0.3623	&	53795	&	57756	&	8.9$\pm$0.06	&	45.52$\pm$0.02	&	45.19$\pm$0.04	&	-1.49$\pm$0.07	&	-1.81$\pm$0.07	&	0.04 $\pm$ 0.02	&	0.07 $\pm$ 0.02	&	0.04 $\pm$ 0.01	&	H$\beta$	&	Turn-off  & CL AGN		&		\\
J114154.47+063509.6	&	175.477	&	6.586	&	0.1018	&	53137	&	59585	&	7.73$\pm$0.03	&	44.26$\pm$0.01	&	43.94$\pm$0.01	&	-1.57$\pm$0.03	&	-1.89$\pm$0.03	&	0.13 $\pm$ 0.03	&	0.12 $\pm$ 0.03	&	0.13 $\pm$ 0.03	&	$\rm{H\beta}$, $\rm{H\alpha}$	&	Turn-off  & CL AGN		&		\\
J114634.92+282642.0	&	176.6455	&	28.445	&	0.2252	&	53799	&	57756	&	7.95$\pm$0.05	&	44.94$\pm$0.02	&	44.79$\pm$0.04	&	-1.11$\pm$0.06	&	-1.27$\pm$0.06	&	0.18 $\pm$ 0.05	&	0.18 $\pm$ 0.05	&	0.18 $\pm$ 0.05	&	H$\beta$	&	Turn-off  & CL AGN		&		\\
J115227.49+320959.5	&	178.1145	&	32.1665	&	0.3743	&	53446	&	57015	&	8.52$\pm$0.1	&	45.43$\pm$0.01	&	44.71$\pm$0.07	&	-1.19$\pm$0.1	&	-1.91$\pm$0.12	&	0.24 $\pm$ 0.06	&	0.14 $\pm$ 0.06	&	0.24 $\pm$ 0.1	&	H$\beta$	&	Turn-off & Candidate	&	(2)	\\
J120332.94+022934.7	&	180.8873	&	2.493	&	0.0772	&	52024	&	56656	&	7.67$\pm$0.08	&	44.77$\pm$0.04	&	44.71$\pm$0.02	&	-1.01$\pm$0.09	&	-1.07$\pm$0.08	&	…	&	…	&	…	&	$\rm{H\beta}$, $\rm{H\alpha}$	&	Turn-off  & Candidate	&		\\
J120447.92+170256.9	&	181.1997	&	17.0491	&	0.2979	&	54207	&	58872	&	8.81$\pm$0.03	&	45.74$\pm$0.01	&	45.18$\pm$0.03	&	-1.17$\pm$0.03	&	-1.72$\pm$0.04	&	0.23 $\pm$ 0.05	&	0.28 $\pm$ 0.05	&	0.23 $\pm$ 0.04	&	H$\beta$	&	Turn-off  & CL AGN		&	(4)	\\
J120635.22+175312.3	&	181.6467	&	17.8868	&	0.1459	&	54207	&	57398	&	8.11$\pm$0.03	&	44.76$\pm$0.01	&	44.81$\pm$0.01	&	-1.45$\pm$0.03	&	-1.4$\pm$0.03	&	0.07 $\pm$ 0.02	&	0.07 $\pm$ 0.02	&	0.07 $\pm$ 0.02	&	H$\beta$	&	Turn-off  & CL AGN		&		\\
J120816.22+300738.3	&	182.0676	&	30.1273	&	0.1384	&	53818	&	57757	&	7.29$\pm$0.01	&	44.12$\pm$0.02	&	42.52$\pm$0.23	&	-1.28$\pm$0.02	&	-2.87$\pm$0.23	&	0.2 $\pm$ 0.03	&	0.09 $\pm$ 0.03	&	0.2 $\pm$ 0.05	&	$\rm{H\beta}$, $\rm{H\alpha}$	&	Turn-off & Candidate	&		\\
J121424.98+252551.5	&	183.6041	&	25.431	&	0.1794	&	54502	&	58490	&	7.29$\pm$0.03	&	43.91$\pm$0.02	&	43.84$\pm$0.1	&	-1.48$\pm$0.04	&	-1.55$\pm$0.1	&	0.16 $\pm$ 0.04	&	0.16 $\pm$ 0.04	&	0.16 $\pm$ 0.04	&	H$\beta$	&	Turn-off & Candidate	&		\\
J122355.23+343856.6	&	185.9801	&	34.6491	&	0.1362	&	53503	&	57052	&	7.62$\pm$0.05	&	44.22$\pm$0.02	&	44.19$\pm$0.02	&	-1.5$\pm$0.05	&	-1.53$\pm$0.05	&	0.07 $\pm$ 0.02	&	0.03 $\pm$ 0.02	&	0.07 $\pm$ 0.04	&	H$\beta$	&	Turn-off  & CL AGN		&		\\
J122623.69+261050.0	&	186.5987	&	26.1806	&	0.1288	&	54502	&	57041	&	8.01$\pm$0.04	&	44.51$\pm$0.01	&	44$\pm$0.09	&	-1.61$\pm$0.04	&	-2.11$\pm$0.09	&	0.28 $\pm$ 0.06	&	0.18 $\pm$ 0.06	&	0.28 $\pm$ 0.1	&	H$\beta$	&	Turn-off  & CL AGN		&		\\
J122945.61+440934.5	&	187.44	&	44.1596	&	0.188	&	53062	&	58492	&	7.91$\pm$0.04	&	44.54$\pm$0.01	&	44.36$\pm$0.01	&	-1.46$\pm$0.04	&	-1.65$\pm$0.04	&	0.06 $\pm$ 0.03	&	0.05 $\pm$ 0.03	&	0.06 $\pm$ 0.02	&	H$\beta$	&	Turn-off  & CL AGN		&		\\
J123020.67+462745.7	&	187.5861	&	46.4627	&	0.1803	&	53084	&	58546	&	8.18$\pm$0.03	&	44.78$\pm$0.01	&	44.32$\pm$0.03	&	-1.49$\pm$0.03	&	-1.96$\pm$0.05	&	0.19 $\pm$ 0.03	&	0.12 $\pm$ 0.03	&	0.19 $\pm$ 0.04	&	H$\beta$	&	Turn-off  & CL AGN		&	(5)	\\
J123359.13+084211.6	&	188.4964	&	8.7032	&	0.2562	&	53474	&	57396	&	7.96$\pm$0.11	&	45.16$\pm$0.03	&	44.38$\pm$0.03	&	-0.9$\pm$0.11	&	-1.68$\pm$0.11	&	0.09 $\pm$ 0.03	&	0.1 $\pm$ 0.03	&	0.09 $\pm$ 0.03	&	H$\beta$	&	Turn-off  & Candidate	& 	(6)	\\
J124554.71+460418.7	&	191.478	&	46.0719	&	0.108	&	53089	&	59637	&	7.46$\pm$0.01	&	44.43$\pm$0.01	&	44.33$\pm$0.06	&	-1.13$\pm$0.01	&	-1.23$\pm$0.06	&	0.07 $\pm$ 0.01	&	0.06 $\pm$ 0.01	&	0.07 $\pm$ 0.01	&	H$\beta$	&	Turn-off  & CL AGN		&		\\
J130339.71+190120.9	&	195.9155	&	19.0225	&	0.0822	&	54499	&	59263	&	…	&	…	&	…	&	…	&	…	&	0.09 $\pm$ 0.01	&	0.05 $\pm$ 0.01	&	0.09 $\pm$ 0.02	&	$\rm{H\beta}$, $\rm{H\alpha}$	&	Turn-on   & Candidate	&		\\
J130501.21+010323.8	&	196.2551	&	1.0566	&	0.1342	&	51986	&	59230	&	7.31$\pm$0.06	&	44.14$\pm$0.04	&	43.61$\pm$0.07	&	-1.28$\pm$0.07	&	-1.81$\pm$0.09	&	0.24 $\pm$ 0.03	&	0.16 $\pm$ 0.03	&	0.24 $\pm$ 0.05	&	$\rm{H\beta}$, $\rm{H\alpha}$	&	Turn-off  & CL AGN		&		\\
J131001.32+143705.2	&	197.5055	&	14.6181	&	0.2134	&	53089	&	57461	&	7.82$\pm$0.07	&	44.5$\pm$0.03	&	44.22$\pm$0.06	&	-1.41$\pm$0.08	&	-1.69$\pm$0.09	&	0.22 $\pm$ 0.08	&	0.27 $\pm$ 0.08	&	0.22 $\pm$ 0.07	&	H$\beta$	&	Turn-off  & CL AGN		&		\\
J131432.70+122718.5	&	198.6363	&	12.4551	&	0.0864	&	53142	&	59996	&	7.66$\pm$0.07	&	44.44$\pm$0.02	&	44$\pm$0.02	&	-1.31$\pm$0.08	&	-1.75$\pm$0.08	&	0.29 $\pm$ 0.04	&	0.19 $\pm$ 0.04	&	0.29 $\pm$ 0.05	&	$\rm{H\beta}$, $\rm{H\alpha}$	&	Turn-off  & CL AGN		&		\\
J131715.12+484630.3	&	199.313	&	48.7751	&	0.1312	&	52759	&	59283	&	6.97$\pm$0.16	&	43.49$\pm$0.06	&	43.48$\pm$0.05	&	-1.57$\pm$0.17	&	-1.59$\pm$0.16	&	0.1 $\pm$ 0.02	&	0.08 $\pm$ 0.02	&	0.1 $\pm$ 0.03	&	H$\beta$	&	Turn-off &Candidate	&		\\
J132421.98+280238.3	&	201.0916	&	28.044	&	0.1244	&	53471	&	56718	&	7.92$\pm$0.06	&	44.52$\pm$0.01	&	44.31$\pm$0.01	&	-1.5$\pm$0.06	&	-1.71$\pm$0.06	&	…	&	…	&	…	&	H$\beta$	&	Turn-off  & CL AGN		&		\\
J133241.18+300106.6	&	203.1716	&	30.0185	&	0.2203	&	53467	&	57778	&	8.28$\pm$0.05	&	44.98$\pm$0.02	&	44.43$\pm$0.04	&	-1.4$\pm$0.05	&	-1.95$\pm$0.06	&	0.17 $\pm$ 0.06	&	0.19 $\pm$ 0.06	&	0.17 $\pm$ 0.05	&	H$\beta$	&	Turn-off  & CL AGN		&	(5)	\\
J133551.98+544450.1	&	203.9666	&	54.7473	&	0.1076	&	52725	&	59641	&	8.07$\pm$0.03	&	44.69$\pm$0.01	&	44.2$\pm$0.03	&	-1.48$\pm$0.03	&	-1.98$\pm$0.04	&	0.15 $\pm$ 0.03	&	0.13 $\pm$ 0.03	&	0.15 $\pm$ 0.03	&	H$\beta$	&	Turn-off  & CL AGN		&		\\
J134133.73+090356.7	&	205.3905	&	9.0658	&	0.105	&	53886	&	59606	&	7.83$\pm$0.07	&	44.86$\pm$0.03	&	44.14$\pm$0.02	&	-1.07$\pm$0.07	&	-1.79$\pm$0.07	&	0.27 $\pm$ 0.04	&	0.2 $\pm$ 0.04	&	0.27 $\pm$ 0.05	&	H$\beta$	&	Turn-off  & CL AGN		&		\\
J134241.02+321231.5	&	205.6709	&	32.2088	&	0.1839	&	53503	&	57105	&	7.35$\pm$0.02	&	44.2$\pm$0.02	&	43.84$\pm$0.08	&	-1.26$\pm$0.02	&	-1.61$\pm$0.08	&	0.36 $\pm$ 0.07	&	0.21 $\pm$ 0.07	&	0.36 $\pm$ 0.12	&	$\rm{H\beta}$, $\rm{H\alpha}$	&	Turn-off & Candidate	&		\\
J134739.65+362240.0	&	206.9152	&	36.3778	&	0.247	&	53476	&	58227	&	8.3$\pm$0.04	&	44.77$\pm$0.01	&	44.09$\pm$0.05	&	-1.62$\pm$0.04	&	-2.31$\pm$0.06	&	0.28 $\pm$ 0.07	&	0.27 $\pm$ 0.07	&	0.28 $\pm$ 0.06	&	$\rm{H\beta}$, $\rm{H\alpha}$	&	Turn-off  & CL AGN		&		\\
J135255.68+252859.6	&	208.232	&	25.4832	&	0.0636	&	53535	&	58553	&	7.4$\pm$0.01	&	44.26$\pm$0	&	43.59$\pm$0.01	&	-1.24$\pm$0.01	&	-1.91$\pm$0.01	&	0.19 $\pm$ 0.04	&	0.16 $\pm$ 0.04	&	0.19 $\pm$ 0.04	&	$\rm{H\beta}$, $\rm{H\alpha}$	&	Turn-off  & CL AGN	&		\\
J135910.91+531933.8	&	209.7955	&	53.3261	&	0.0753	&	52468	&	58137	&	8.09$\pm$0.01	&	43.94$\pm$0.01	&	43.58$\pm$0.02	&	-2.25$\pm$0.01	&	-2.6$\pm$0.03	&	0.02 $\pm$ 0.02	&	0.01 $\pm$ 0.02	&	0.02 $\pm$ 0.03	&	$\rm{H\beta}$, $\rm{H\alpha}$	&	Turn-off  & Candidate	&		\\
J140207.59+284859.3	&	210.5316	&	28.8165	&	0.22	&	53852	&	57815	&	8.26$\pm$0.04	&	44.66$\pm$0.01	&	44.27$\pm$0.02	&	-1.7$\pm$0.04	&	-2.09$\pm$0.05	&	0.16 $\pm$ 0.03	&	0.09 $\pm$ 0.03	&	0.16 $\pm$ 0.04	&	$\rm{H\beta}$, $\rm{H\alpha}$	&	Turn-off  & CL AGN	&		\\
J142336.99+103812.4	&	215.9041	&	10.6368	&	0.1877	&	53885	&	60029	&	7.32$\pm$0.03	&	44.16$\pm$0.03	&	43.81$\pm$0.07	&	-1.25$\pm$0.04	&	-1.6$\pm$0.08	&	0.16 $\pm$ 0.03	&	0.14 $\pm$ 0.03	&	0.16 $\pm$ 0.03	&	H$\beta$	&	Turn-off & Candidate	&		\\
J143426.20+054201.0	&	218.6092	&	5.7003	&	0.1572	&	53504	&	58931	&	8.06$\pm$0.04	&	44.25$\pm$0.01	&	43.6$\pm$0.05	&	-1.9$\pm$0.04	&	-2.56$\pm$0.06	&	0.07 $\pm$ 0.01	&	0.04 $\pm$ 0.01	&	0.07 $\pm$ 0.02	&	H$\alpha$	&	Turn-off & Candidate	&		\\
J144505.08+524341.9	&	221.2712	&	52.7283	&	0.3246	&	52781	&	57461	&	8.38$\pm$0.03	&	45.22$\pm$0.01	&	44.78$\pm$0.05	&	-1.26$\pm$0.03	&	-1.7$\pm$0.06	&	0.08 $\pm$ 0.03	&	0.04 $\pm$ 0.03	&	0.08 $\pm$ 0.03	&	H$\beta$	&	Turn-off & Candidate	&		\\
J145359.73+091543.4	&	223.4989	&	9.262	&	0.279	&	53827	&	58897	&	8.03$\pm$0.09	&	45.01$\pm$0.06	&	44.7$\pm$0.01	&	-1.12$\pm$0.11	&	-1.43$\pm$0.1	&	0.16 $\pm$ 0.05	&	0.17 $\pm$ 0.05	&	0.16 $\pm$ 0.06	&	H$\beta$	&	Turn-off  & CL AGN	&		\\
J150026.17+374732.4	&	225.109	&	37.7923	&	0.1806	&	52790	&	57844	&	7.58$\pm$0.04	&	44.52$\pm$0.02	&	44.21$\pm$0.03	&	-1.16$\pm$0.05	&	-1.47$\pm$0.05	&	0.09 $\pm$ 0.02	&	0.05 $\pm$ 0.02	&	0.09 $\pm$ 0.03	&	H$\beta$	&	Turn-off  & CL AGN		&		\\
J150238.37+241313.8	&	225.6599	&	24.2205	&	0.0637	&	54509	&	59702	&	7.52$\pm$0.03	&	43.87$\pm$0.01	&	43.54$\pm$0.01	&	-1.75$\pm$0.03	&	-2.08$\pm$0.03	&	0.22 $\pm$ 0.03	&	0.12 $\pm$ 0.03	&	0.22 $\pm$ 0.04	&	H$\alpha$	&	Turn-off  & CL AGN	&		\\
J151143.41+210104.0	&	227.9309	&	21.0178	&	0.0805	&	54525	&	59732	&	8.15$\pm$0.03	&	44.21$\pm$0.02	&	44.85$\pm$0	&	-2.04$\pm$0.04	&	-1.4$\pm$0.03	&	0.33 $\pm$ 0.06	&	0.31 $\pm$ 0.06	&	0.33 $\pm$ 0.06	&	H$\beta$	&	Turn-on	   & CL AGN	&		\\
J151219.33+293737.0	&	228.0806	&	29.6269	&	0.1999	&	54144	&	57106	&	8.09$\pm$0.05	&	44.76$\pm$0.01	&	44.59$\pm$0.01	&	-1.43$\pm$0.05	&	-1.6$\pm$0.05	&	…	&	…	&	…	&	H$\beta$	&	Turn-off  & CL AGN	&		\\
J151804.33+364747.9	&	229.518	&	36.7966	&	0.1971	&	53470	&	57458	&	7.67$\pm$0.07	&	44.55$\pm$0.02	&	44.53$\pm$0.02	&	-1.22$\pm$0.08	&	-1.24$\pm$0.07	&	0.1 $\pm$ 0.06	&	0.06 $\pm$ 0.06	&	0.1 $\pm$ 0.03	&	H$\beta$	&	Turn-off  & CL AGN	&		\\
J152134.96+320638.6	&	230.3957	&	32.1107	&	0.1117	&	53118	&	57134	&	7.51$\pm$0.04	&	44$\pm$0.02	&	43.92$\pm$0.02	&	-1.61$\pm$0.05	&	-1.69$\pm$0.05	&	0.08 $\pm$ 0.03	&	0.06 $\pm$ 0.03	&	0.08 $\pm$ 0.03	&	$\rm{H\beta}$, $\rm{H\alpha}$	&	Turn-off & Candidate &		\\
J152904.77+024640.4	&	232.2699	&	2.7779	&	0.0974	&	52026	&	57834	&	7.61$\pm$0.06	&	44.3$\pm$0.04	&	44.24$\pm$0.03	&	-1.42$\pm$0.07	&	-1.47$\pm$0.07	&	0.08 $\pm$ 0.04	&	0.08 $\pm$ 0.04	&	0.08 $\pm$ 0.03	&	H$\beta$	&	Turn-off  & CL AGN		&		\\
J153508.49+131730.5	&	233.7854	&	13.2918	&	0.0935	&	54265	&	57869	&	7.74$\pm$0.04	&	44.09$\pm$0.01	&	44.23$\pm$0.04	&	-1.75$\pm$0.04	&	-1.6$\pm$0.05	&	0.05 $\pm$ 0.02	&	0.04 $\pm$ 0.02	&	0.05 $\pm$ 0.02	&	H$\alpha$	&	Turn-off & Candidate	&		\\
J155105.31+261819.2	&	237.7721	&	26.3053	&	0.2415	&	53498	&	57110	&	7.99$\pm$0.04	&	44.83$\pm$0.01	&	44.17$\pm$0.03	&	-1.26$\pm$0.04	&	-1.92$\pm$0.05	&	0.18 $\pm$ 0.02	&	0.07 $\pm$ 0.02	&	0.18 $\pm$ 0.06	&	H$\beta$	&	Turn-off & Candidate	&		\\
J160157.53+490207.8	&	240.4897	&	49.0355	&	0.2465	&	52054	&	59674	&	8.45$\pm$0.02	&	45.17$\pm$0.01	&	44.53$\pm$0.02	&	-1.38$\pm$0.02	&	-2.02$\pm$0.03	&	0.16 $\pm$ 0.03	&	0.15 $\pm$ 0.03	&	0.16 $\pm$ 0.03	&	$\rm{H\beta}$, $\rm{H\alpha}$	&	Turn-off  & CL AGN		&		\\
J161711.42+063833.5	&	244.2976	&	6.6426	&	0.2291	&	53501	&	56776	&	8.42$\pm$0.01	&	45.55$\pm$0	&	45.07$\pm$0.02	&	-0.97$\pm$0.01	&	-1.45$\pm$0.02	&	…	&	…	&	…	&	H$\beta$	&	Turn-off  & CL AGN		&	(6)	\\
J171617.23+273410.1	&	259.0718	&	27.5695	&	0.2888	&	52431	&	57865	&	7.95$\pm$0.05	&	44.61$\pm$0.02	&	44.44$\pm$0.01	&	-1.44$\pm$0.05	&	-1.61$\pm$0.05	&	0.11 $\pm$ 0.02	&	0.09 $\pm$ 0.02	&	0.11 $\pm$ 0.03	&	H$\beta$	&	Turn-off & Candidate	&		\\
J172533.07+571645.6	&	261.3878	&	57.2793	&	0.0659	&	51818	&	58220	&	7.8$\pm$0.01	&	44.03$\pm$0.01	&	43.26$\pm$0.06	&	-1.87$\pm$0.01	&	-2.64$\pm$0.06	&	0.1 $\pm$ 0.03	&	0.07 $\pm$ 0.03	&	0.1 $\pm$ 0.03	&	H$\alpha$	&	Turn-off & Candidate	&		\\
J220701.85+041706.2	&	331.7577	&	4.2851	&	0.0603	&	55480	&	57307	&	7.22$\pm$0.03	&	43.53$\pm$0.01	&	42.95$\pm$0.04	&	-1.79$\pm$0.04	&	-2.37$\pm$0.05	&	0.02 $\pm$ 0.02	&	0.02 $\pm$ 0.02	&	0.02 $\pm$ 0.07	&	H$\beta$	&	Turn-off & Candidate	&		\\
J223716.91+142432.8	&	339.3205	&	14.4091	&	0.1857	&	52520	&	57327	&	7.8$\pm$0.1	&	44.22$\pm$0.01	&	43.59$\pm$0.1	&	-1.68$\pm$0.1	&	-2.32$\pm$0.14	&	0.07 $\pm$ 0.01	&	0.03 $\pm$ 0.01	&	0.07 $\pm$ 0.03	&	H$\beta$	&	Turn-off  & Candidate	&		\\
\enddata
\tablenotetext{  }{NOTE—Columns: (1) object name, (2) right ascension, (3) declination, (4) redshift, (5) MJD for the first epoch, (6) MJD for the second epoch, (7) black hole mass, (8) the bolometric luminosity calculated from the first epoch, (9) the bolometric luminosity calculated from the second epoch, (10) the Eddington rate of the first epoch, (11) the Eddington rate of the second epoch, (12) the intrinsic variability amplitude of r-band, (13) the intrinsic variability amplitude of W1 light curves, (14) the intrinsic variability amplitude of W2 light curves, (15) the emission lines that have the notable changes for the repeated observations, (16) the transition type, (17) CL AGNs are sources exhibiting disappearance or emergence of \hb\ emission line, along with either detectable broad \ha\ line or significant photometric variability; candidates are sources exhibiting the disappearance or emergence of both \hb\ and \ha\ emission line, while without significant photometric variability, (18) the CL~AGNs has reported by previous study: (1) \citet{2022ApJ...933..180G}; (2) \citet{yang2018discovery}; (3) \citet{2016ApJ...821...33R}; (4) \citet{2019ApJ...887...15W}; (5) \citet{2024ApJ...966..128W}; (6) \citet{2019ApJ...874....8M}. }
\label{CLAGNs}
\end{deluxetable}
\end{longrotatetable}

\bibliography{ref}{}
\bibliographystyle{aasjournal}

\end{document}